\documentclass{blois}

\usepackage{amsmath}
\usepackage{amssymb}
\newcommand{\fb}{\,\text{fb}}
\newcommand{\invfb}{\,\text{fb}^{-1}}
\newcommand{\eV}{\,\text{eV}}
\newcommand{\MeV}{\,\text{MeV}}
\newcommand{\GeV}{\,\text{GeV}}
\newcommand{\TeV}{\,\text{TeV}}

\def\be{\begin{equation}}
\def\ee{\end{equation}}
\def\bea{\begin{eqnarray}}
\def\eea{\end{eqnarray}}

\newcommand{\Photo}{}

\begin{document}
\begin{flushright}
  CERN-PH-TH-2015-234\vspace{-0.5cm}
\end{flushright}
\vspace*{4cm}
\title{Summary of the XXVIIth Rencontres de Blois: Particle Physics
  and Cosmology}

\author{GAVIN P. SALAM\,\footnote{On leave from CNRS, UMR 7589, LPTHE,
    F-75005, Paris, France.}}

\address{CERN, PH-TH, CH-1211 Geneva 23, Switzerland}

\maketitle\abstracts{
  This writeup summarises some of the highlights from the 2015
  Rencontres de Blois, with a compression ratio of about 100:1
  relative to the original presentations.
}

\section{Introduction}

The XXVIIth Rencontres de Blois has taken place at a special moment in
particle physics and cosmology, one where it is a privilege to take
stock of the shape of the fields of particle physics and cosmology:
nearly all results from Run 1 of the Large Hadron Collider have now
emerged, the Planck experiment has released many of its main
findings and there is also a wealth of data from cosmic-ray
experiments.
At the same time, we can look forward to Run 2 of the LHC, which
started as the conference was taking place, and much progress in the
near future also in dark matter searches.

This summary selects a few of the highlights from the roughly 130
talks of the conference (mostly the 31 plenary talks), with the
perspective of a particle physicist, but attempting to reflect the
roughly 1:1 balance of LHC and non-LHC subjects.
The selection is woefully incomplete, and the reader is referred to
the complete proceedings for the details of the many subjects that
were touched upon.

\section{Higgs, top and other standard-model physics}

The widely celebrated major achievement of Run~I of the LHC was
the discovery of the Higgs boson. 
The Higgs boson comes late to the stable of standard-model particles.
At least in part, this is a consequence of its rather low production
cross section, which is an order of magnitude smaller, say, than the
$t\bar t$ cross section in $8$\,TeV proton-proton collisions at the LHC.

\begin{figure}
  \centering
  \includegraphics[width=0.5\textwidth]{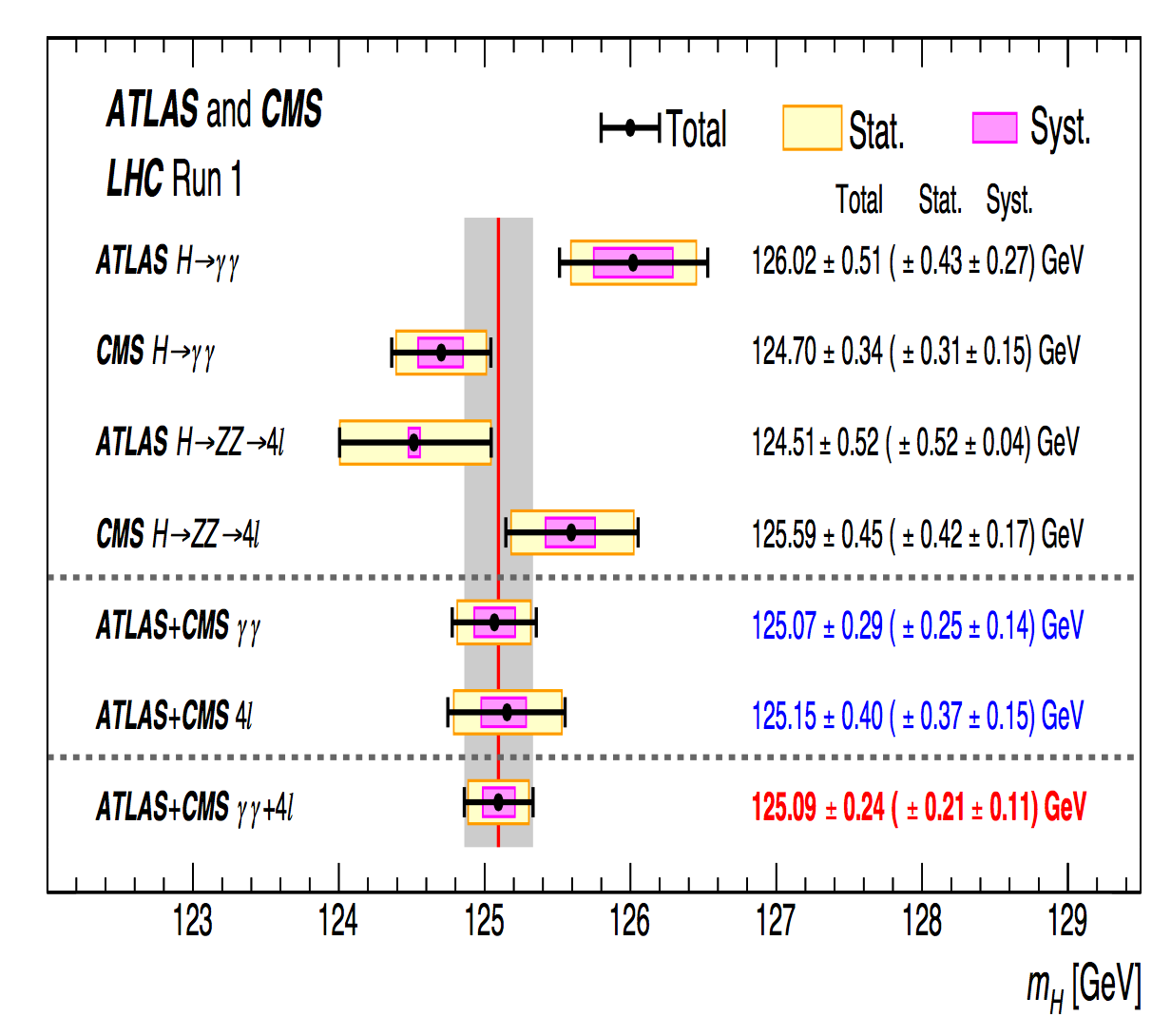}\hfill
  \includegraphics[width=0.45\textwidth]{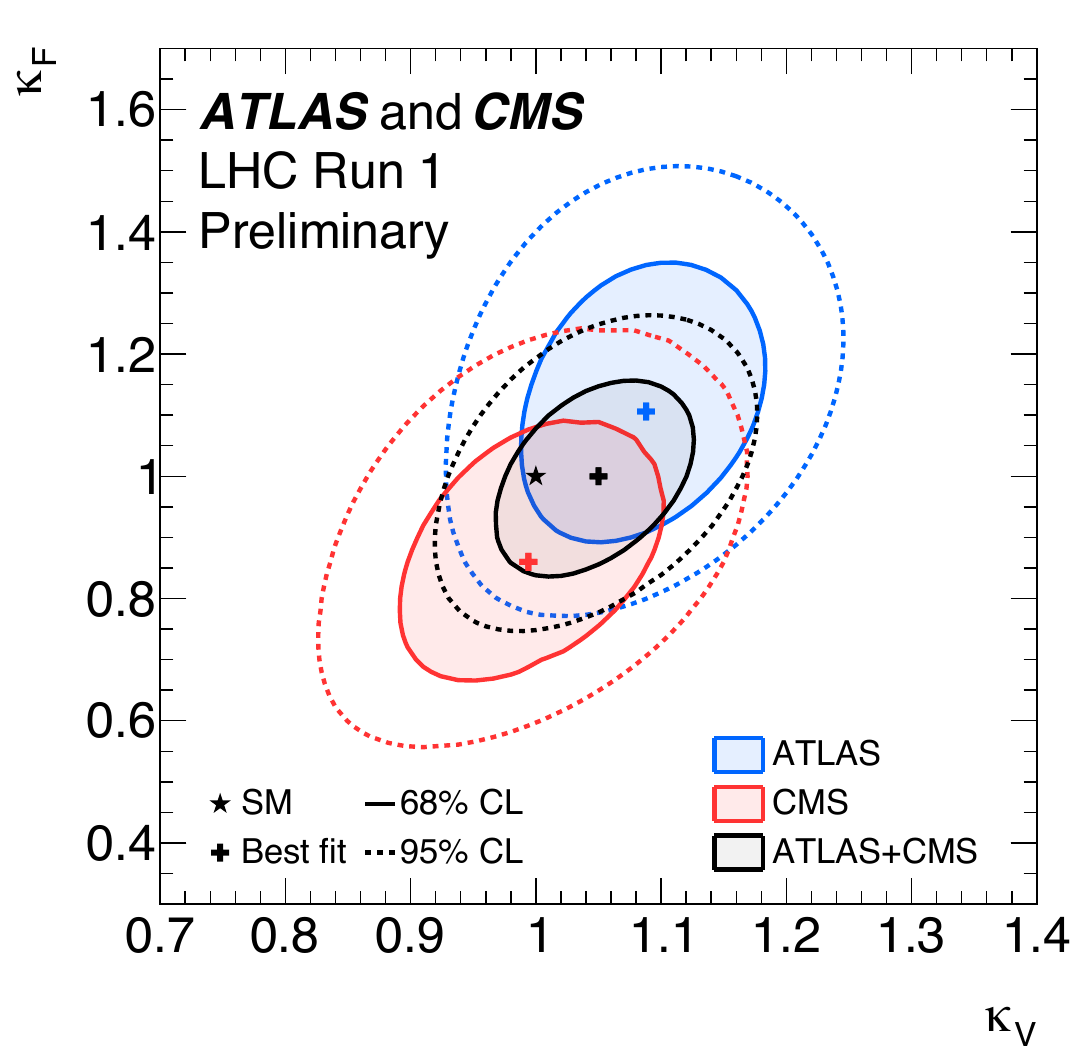}
  \caption{Left: combined ATLAS and CMS Higgs boson mass measurements (figure from the talk by
    Gomez\,\protect\cite{GomezProceedings}).
    Right: joint ATLAS and CMS fit for the scaling factors $\kappa_F$
    and $\kappa_V$ that multiply the fermionic and vector couplings of the
    Higgs boson.\,\protect\cite{ATLASCMSHiggsFits} 
  }
  \label{fig:Higgs-mass}
\end{figure}

We have, however, been exceptionally lucky with the mass of the Higgs
boson. 
At 125\,GeV, a wide range of its decays have already proved amenable to
some degree of study, including those to $\gamma \gamma$, $Z Z^*$,
$WW*$, $\tau^+ \tau^-$, $b\bar b$.
Thanks to the decays to $\gamma \gamma$ and $Z Z^*$, it is possible to
measure its mass precisely: though barely 3-years have passed since
discovery, the $0.2\%$ precision obtained on the Higgs
mass,\cite{PetersProceedings,GomezProceedings}
Fig.~\ref{fig:Higgs-mass} (left), already surpasses the 
roughly $0.4\%$ that is typically quoted for the top-quark
mass~\cite{ListerProceedings}.
The data are also clearly consistent with a $0^+$ spin--parity assignment. 

\begin{figure}
  \centering
  \begin{minipage}{0.45\linewidth}
    \includegraphics[width=\textwidth]{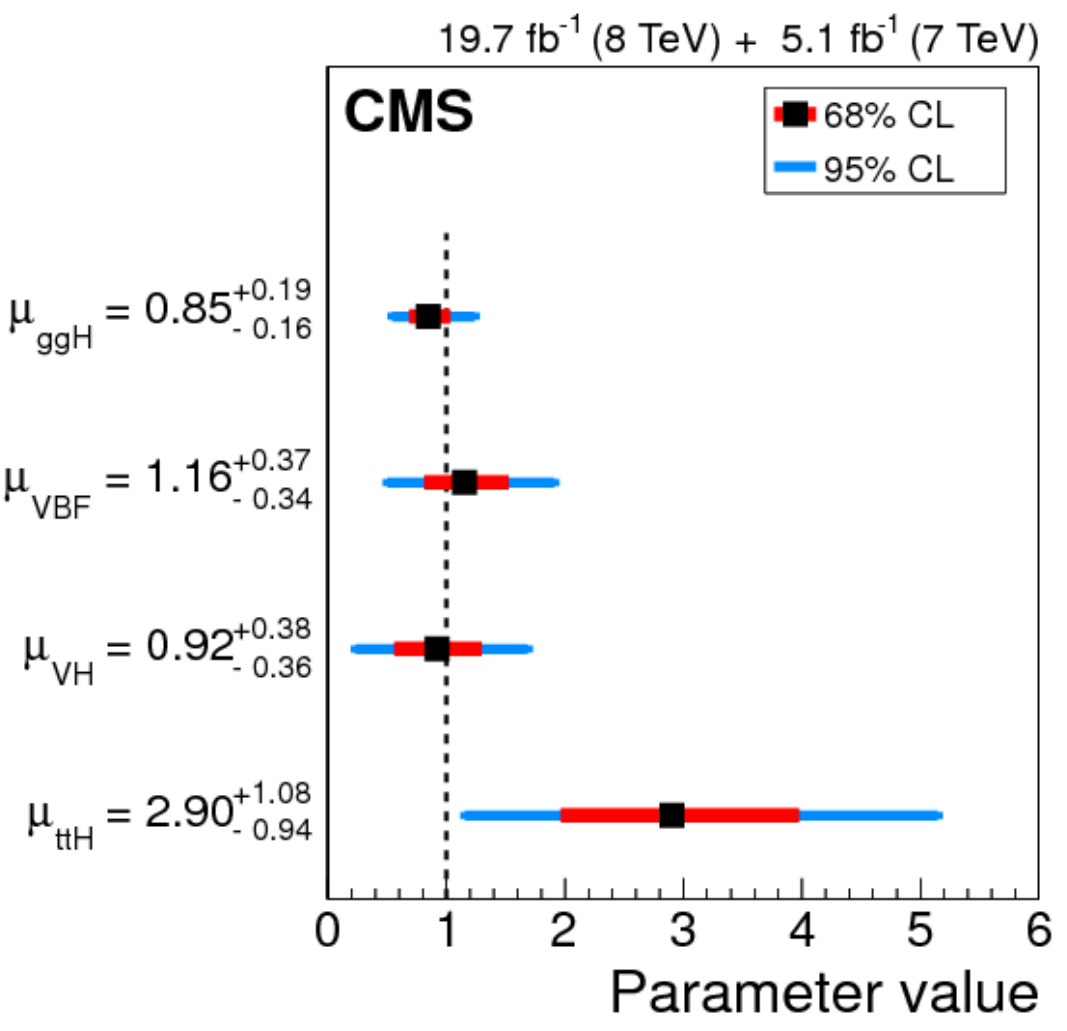}
  \end{minipage}\hfill
  \begin{minipage}{0.53\linewidth}
    \includegraphics[width=\textwidth]{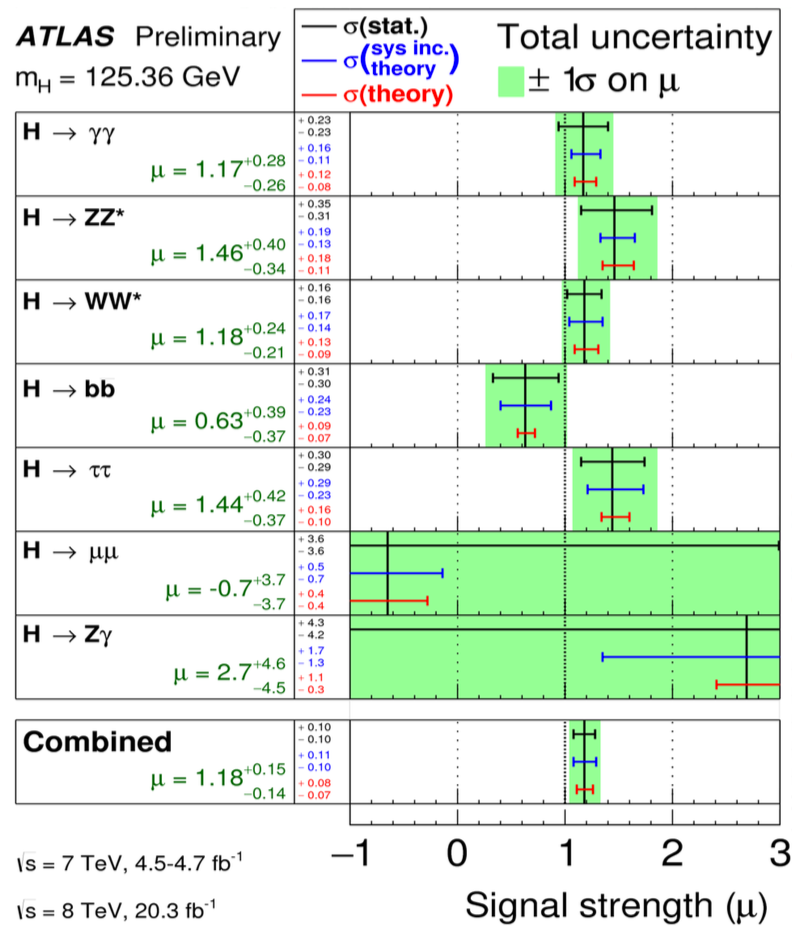}
  \end{minipage}
  \caption{Left: best signal-strength values, $\mu \equiv
    \sigma/\sigma_{\text{SM}}$, from CMS for different Higgs-boson
    production channels
    Right:  best signal-strength values from ATLAS for different
    decays of the Higgs boson.
    Figures from talks by 
    Gomez and Peters.\protect\cite{GomezProceedings,PetersProceedings}
    }
  \label{fig:Higgs-prodn-decay}
\end{figure}

As illustrated in Fig.~\ref{fig:Higgs-prodn-decay}, the overall rate
of Higgs production has been constrained at the level of about $15\%$ and
a number of its branching ratios have been determined with
precisions in the range of $20-40\%$ (modulo an overall
normalisation).
There are even first constraints on its total width.
The data have been analysed in myriad other ways, one of the most
common being the extraction of ``$\kappa$-factors'': e.g.\ one allows
for a rescaling of all vector couplings by a factor $\kappa_V$ and all
fermion couplings by $\kappa_F$ and attempts to determine the allowed
range of $\kappa_V,\kappa_F$ values.
For both ATLAS and CMS, there is excellent agreement with the
standard-model expectation of $\kappa_F = \kappa_V = 1$, to within
$10-20\%$ (during the conference only separate ATLAS and CMS fits were
available; since then a joint fit has appeared and it is this that is
shown in Fig.~\ref{fig:Higgs-mass} (right)).
Fits of this kind have been carried out using also electroweak
precision data. 
Within certain assumptions as to how a change of $\kappa_V$ would
affect the EW $S$ and $T$ parameters, this would bring further
significant reduction in the $\kappa_V$ uncertainty, to a few
percent.\,\cite{PeifferProceedings}

Going forwards, the clear path for the LHC is towards significantly
greater precision. 
Over the course of the next run, the LHC experiments should produce 10
times more Higgs bosons.
The resulting factor of three improvement in precision that can be
expected from this larger dataset will require significant advances
also in our theoretical treatment and predictions.
For example, there is much discussion about replacing or supplementing
the ``$\kappa$-framework'' with an effective field theory treatment of
possible deviations from the standard model~\cite{MassoProceedings}.
This would, for example, make it more straightforward to consistently
incorporate (large) higher-order QCD corrections in the analysis.

\begin{figure}
  \centering
  \begin{minipage}{0.43\linewidth}
    \includegraphics[width=\textwidth]{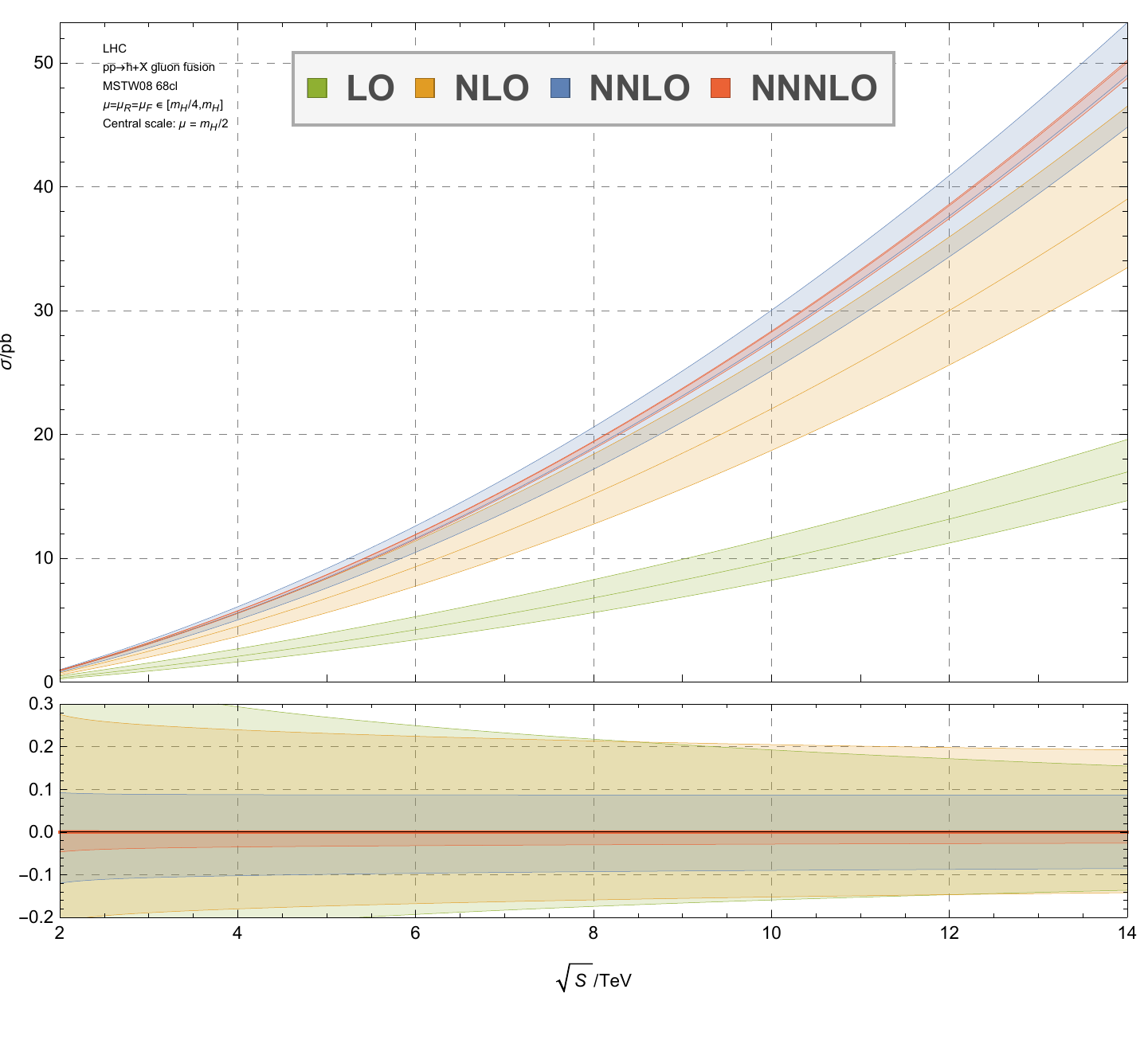}
  \end{minipage}\hfill
  \begin{minipage}{0.54\linewidth}
    \includegraphics[width=\textwidth]{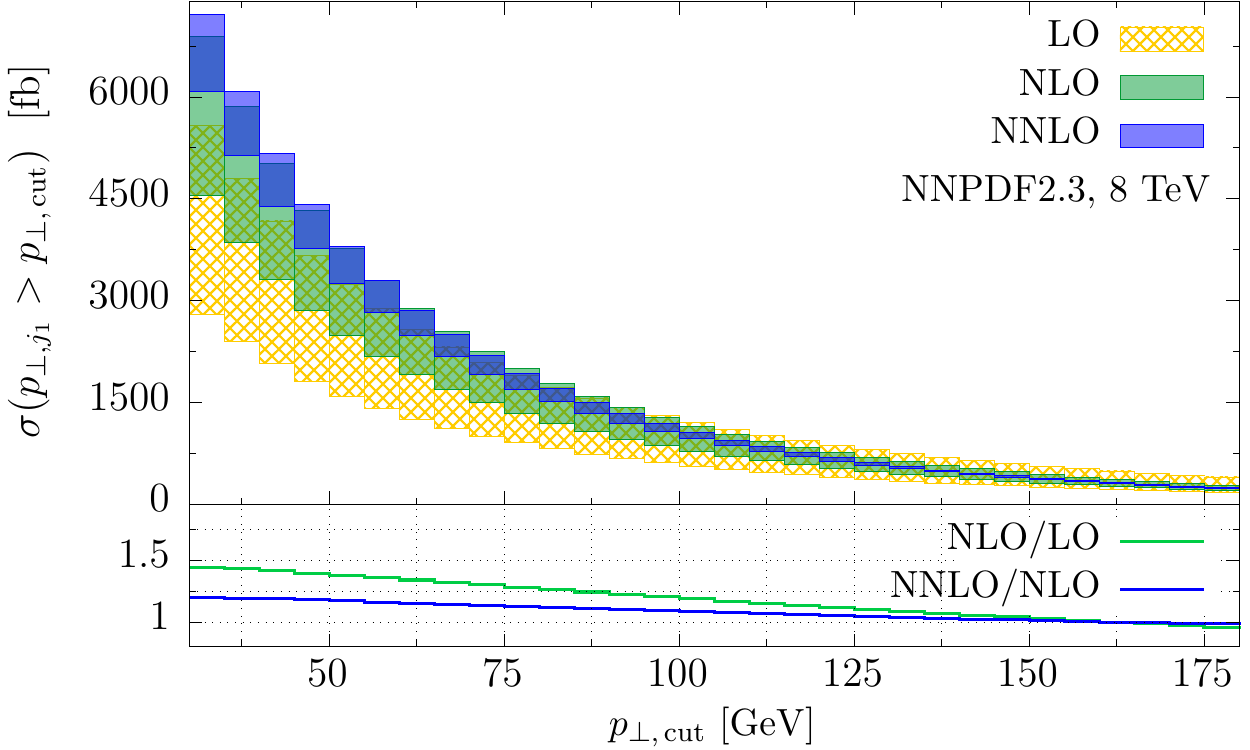}
  \end{minipage}
  \caption{Left: the Higgs boson total cross section in gluon fusion,
    as a function of the centre-of-mass energy $\sqrt{s}$, including
    the latest NNNLO prediction with its considerably reduced scale
    uncertainty band (figure\,\protect\cite{Anastasiou:2015ema} as
    shown in the talk by
    Mistlberger\,\protect\cite{MistlbergerProceedings}).
    Right: the cross section for a Higgs boson to be produced in
    association with a jet, as a function of the jet transverse
    momentum threshold, $p_{\perp,\text{cut}}$
    (figure\,\protect\cite{Boughezal:2015dra} as shown in the talk by
    Caola\,\protect\cite{CaolaProceedings}).  }
  \label{fig:QCD-calculations}
\end{figure}

Within the standard-model framework, the large QCD corrections
are already critical: the state of the art for the total cross section
was until recently next-next-to-leading order (NNLO) in perturbation
theory, leading to theory uncertainties from missing higher orders of
about $7-9\%$.
Such uncertainties are comparable with today's experimental systematic
and statistical errors and so would become the limiting factor for
future analyses.
Fortunately, there has in recent months been very considerable
progress on theoretical calculations: the perturbative uncertainty on
the largest Higgs production process, gluon-fusion (in a large
top-mass approximation) has been reduced down to about $3\%$ thanks to
the first ever NNNLO QCD calculation for a hadron collider
process.\,\cite{MistlbergerProceedings}
Calculations of differential cross sections, notably for Higgs
production with an additional jet, have also seen significant
advances, and one such result was also presented at this
workshop.\,\cite{CaolaProceedings}
Plots illustrating these results are shown in Fig.~\ref{fig:QCD-calculations}.
A number of other calculations of similar complexity have also become
available in the past few months. 
%

\begin{figure}
  \centering
  \begin{minipage}{0.38\linewidth}
    \includegraphics[width=\textwidth]{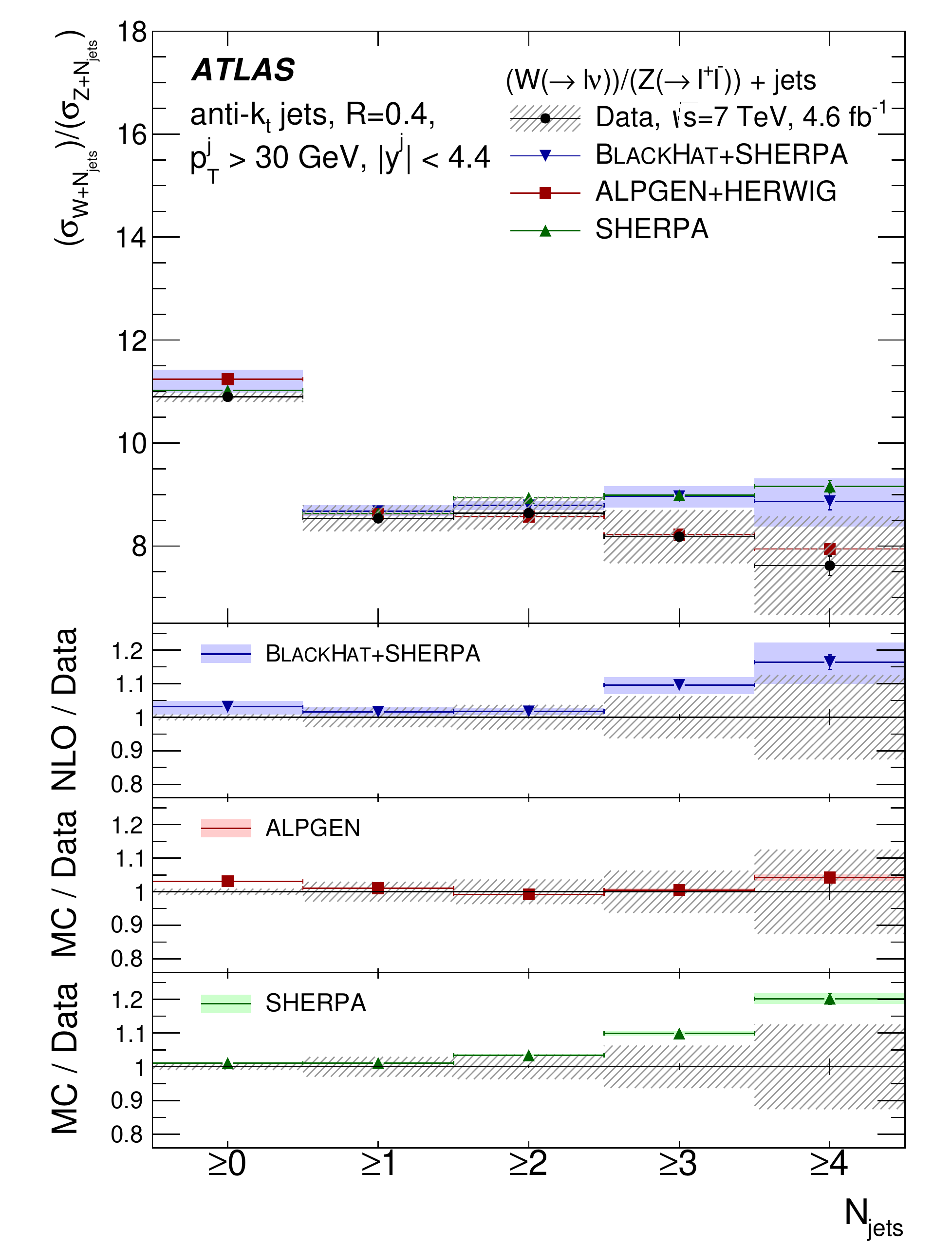}
  \end{minipage}\hfill
  \begin{minipage}{0.58\linewidth}
    \includegraphics[width=\textwidth]{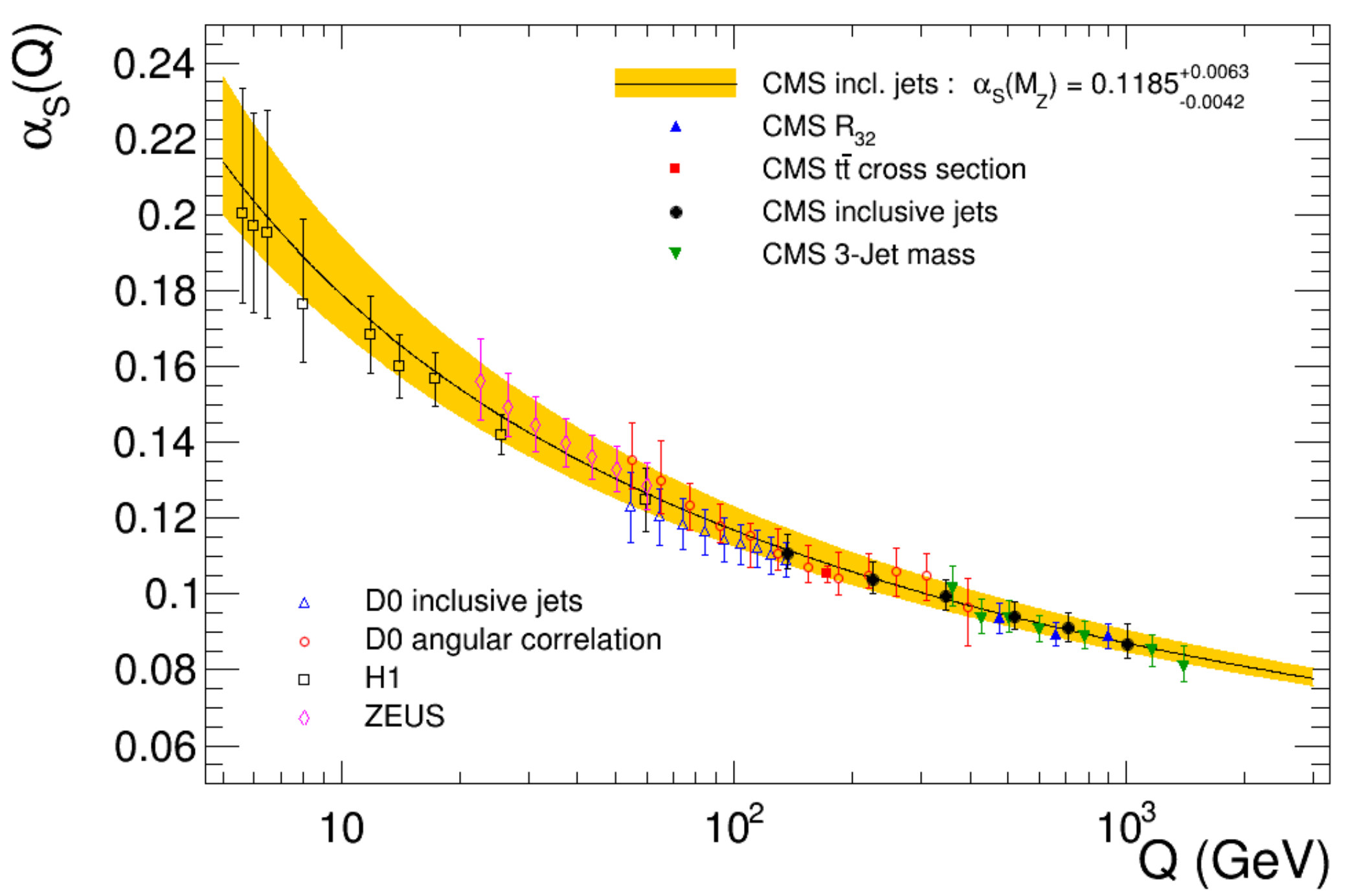}
  \end{minipage}
  \caption{Left: illustration of the comparison between theory
    predictions and ATLAS data for the W and Z cross section ratios,
    as a function of the number of jets
    (figure\,\protect\cite{Aad:2014rta} as shown in the talk by
    Savin\,\protect\cite{SavinProceedings}).
    Right: extractions of the strong coupling over a wide range of scales $Q$, from
    CMS and other experiments (taken from the talk by
    Savin\,\protect\cite{SavinProceedings}).
  }
  \label{fig:LHC-data-v-theory}
\end{figure}

Interpretations and predictions for hadron colliders involve far more
than QCD perturbative calculations, as discussed by
Dittmaier:\,\cite{DittmaierProceedings} electroweak corrections can be
crucial, as can resummations of logarithmically enhanced contributions
(e.g.\ for processes with multiple scales), knowledge of parton
distributions (cf.\ Ref.~\cite{BertoneProceedings}), and more-or-less
controlled modelling of non-perturbative QCD effects.\footnote{In some
  cases one may also gain insight into non-perturbative effects
  through dualities; such methods were discussed by
  Son\,\protect\cite{SonProceedings} in the context of condensed
  matter physics, where they have seen extensive study in the past few
  years.}
A crucial and very active part of the LHC programme is to test and
further understand this rich panoply of
physics.\,\cite{SavinProceedings}
While agreement is usually good (Fig.~\ref{fig:LHC-data-v-theory}),
there are some places with moderate disagreement, typically at the
10--20\% level.
Further study of these regions of disagreement will undoubtedly help
drive progress in our understanding hadron-collider physics.

Among the range of physics processes being studied at the LHC, those
involving top quarks have a special place.\cite{ListerProceedings}
For example, the top is unique in having a Yukawa coupling close to 1,
and it decays before hadronisation.
Various scenarios of new physics assign a special role to the top
quark itself or to partners of the top quark (e.g.\ a stop squark) and
top-quarks then inevitably find their way into new-particle decays.
More annoyingly, standard-model top production is a major background
to many processes of interest.

\begin{figure}
  \centering
  \begin{minipage}{0.48\linewidth}
    \includegraphics[width=\textwidth]{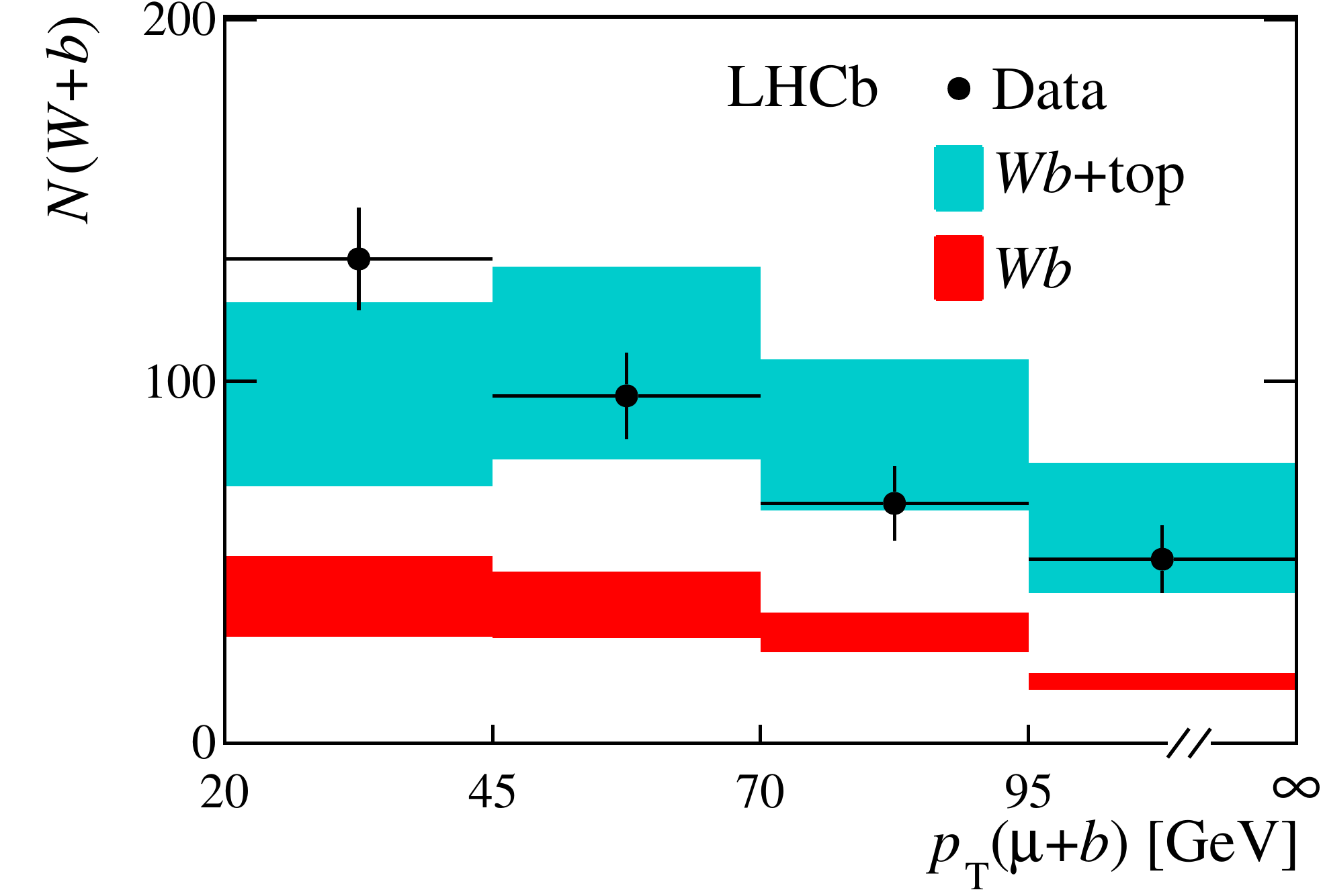}
  \end{minipage}\hfill
  \begin{minipage}{0.48\linewidth}
    \includegraphics[width=\textwidth]{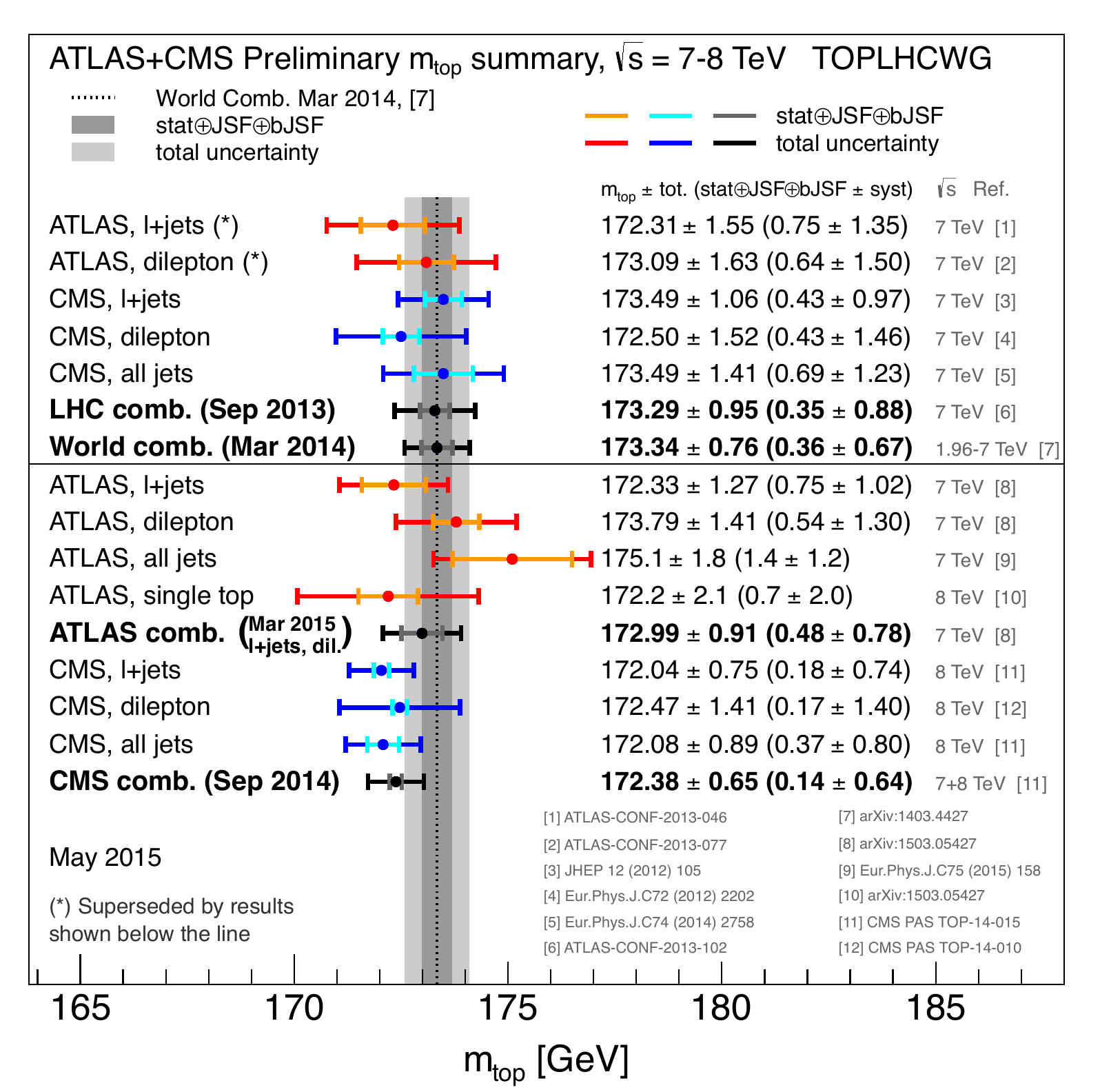}
  \end{minipage}\hfill
  \caption{Left: results from the LHCb experiment showing observation
    of top quarks (figure as shown by
    Barter\,\protect\cite{BarterProceedings,Aaij:2015mwa}).
    Right: summary of top-mass measurements at the LHC (figure shown by
    Lister\,\protect\cite{ListerProceedings}).
  }
  \label{fig:top-results}
\end{figure}

Until recently only four experiments had ever observed top-quark
production: CDF, D0, ATLAS and CMS. 
This Rencontres de Blois saw the first conference presentation of
results by the LHCb collaboration, showing that they too have now
joined the exclusive club of experiments that can study the top, with
observation at just over $5\sigma$
significance,\cite{BarterProceedings} cf.\ Fig.~\ref{fig:top-results} (left).
LHCb studies a complementary kinematic regime relative to ATLAS and
CMS, forward instead of central, and this complementarity is likely to
considerably enrich LHC top studies in the years to come.

I commented above on the fact that our knowledge of the Higgs mass
already surpasses that of the top mass, cf.\
Fig.~\ref{fig:top-results} (right), even though top quarks have
been studied for the past 20 years (and Higgs couplings are known with
an accuracy approaching that of the main top ``coupling'', $V_{tb}$).
There are many reasons for this, including the fact that the top quark
has no purely photonic or leptonic decays, and the complications
associated with the fact that it is a coloured object that decays
before it hadronises.
However, thanks to its (relatively) large production cross section,
there are other aspects where top-quark studies clearly surpass
today's Higgs studies.
This is especially the case for differential distributions, which
extend up to the TeV scale, both in ``standard-model'' studies (which
show reasonable consistency, albeit with some tension for the the
top-quark transverse momentum distribution) and in searches, with
sensitivity to new resonances approaching 2~TeV.

\section{LHC new-physics searches}

\begin{figure}
  \centering
  \begin{minipage}{0.43\linewidth}
    \includegraphics[width=\textwidth]{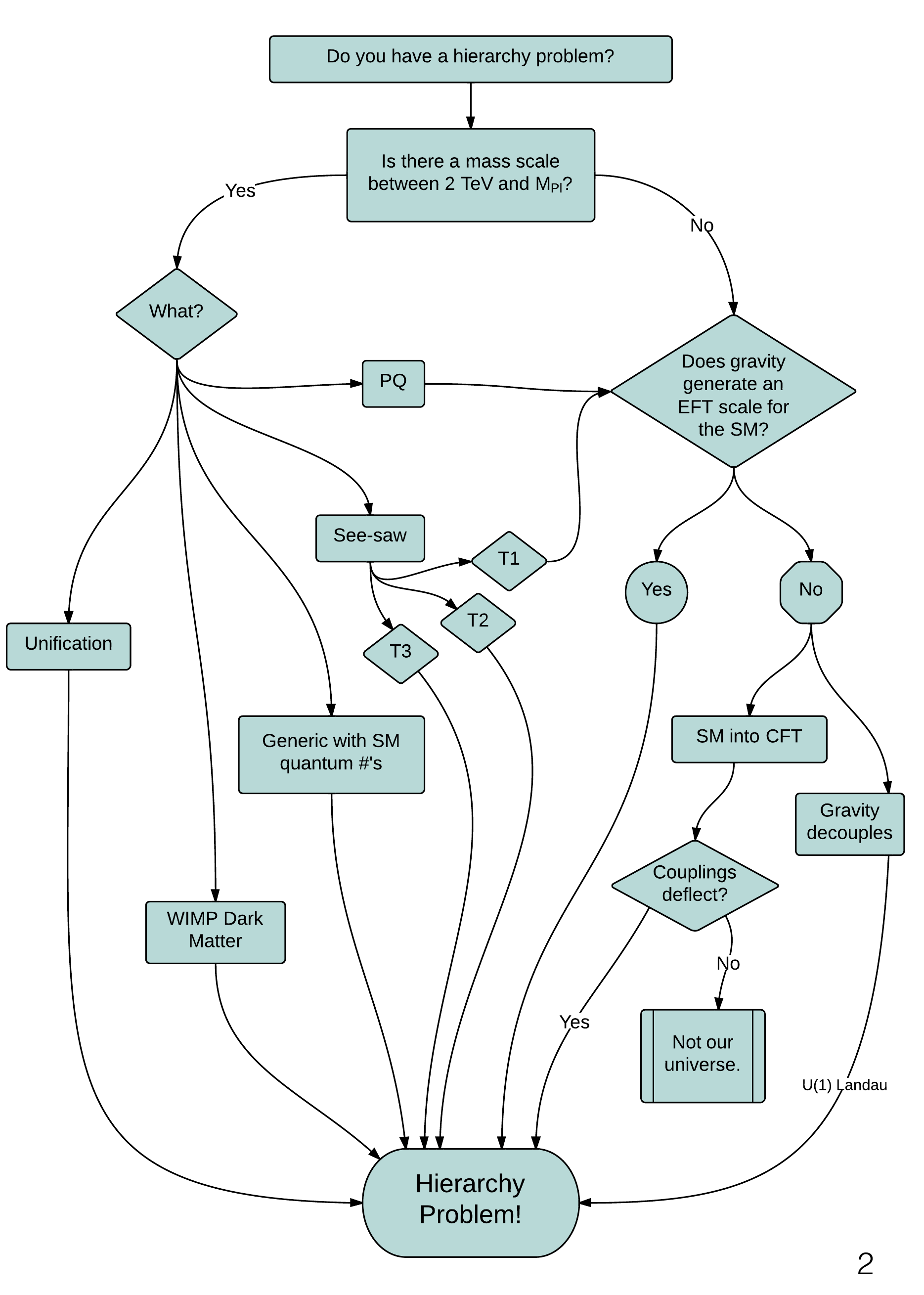}
  \end{minipage}\hfill
  \begin{minipage}{0.53\linewidth}
    \includegraphics[width=\textwidth]{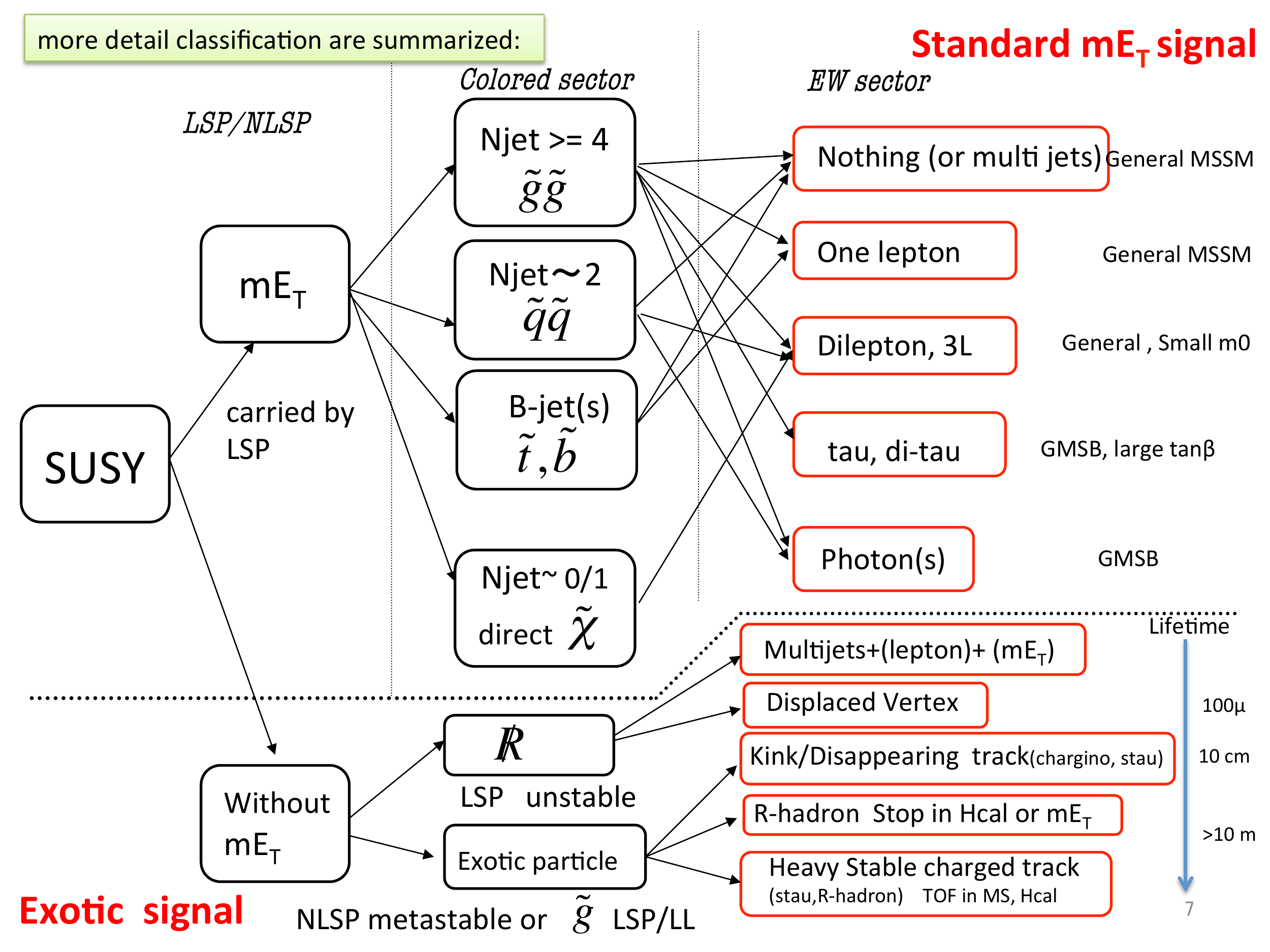}
  \end{minipage}
  \caption{Left: flowchart illustrating difficulties in evading the
    hierarchy problem without new physics near the electroweak scale
    (taken from talk by Craig\,\protect\cite{CraigProceedings}).
    Right: breakdown of different search strategies for supersymmetry
    at the LHC (taken from talk by Asai\,\protect\cite{AsaiProceedings}).
  }
  \label{fig:susy-motivations-and-searches}
\end{figure}

This naturally brings us to the question of new-physics searches. 
It is widely believed that elementary scalars are quadratically
sensitive to physics at higher scales.
Gravity attests to the presence of a higher scale and it seems
difficult to ensure that a fundamental massless scalar like the Higgs
remains at the electroweak scale without the presence of some new
physics nearby,\footnote{Or nearly thirty orders of magnitude of fine
  tuning, possibly anthropically driven.} as illustrated also in a
flowchart, Fig.~\ref{fig:susy-motivations-and-searches}
(left), by Craig.\cite{CraigProceedings}

The most extensively studied candidate for physics beyond the standard
model is undoubtedly supersymmetry (SUSY), with
Fig.~\ref{fig:susy-motivations-and-searches} (right) illustrating the
many topologies in which it might be discovered.
As emphasised by various speakers, there are many arguments in its
favour such as naturalness, the fact that it provides a candidate for
dark matter, unification of the couplings, consistency with the Higgs
mass and so forth.
Yet, it is common to hear that simplest versions are now excluded up
to a mass scale of about $1.5$\,TeV, which in itself brings a fine
tuning of about $1\%$.

\begin{figure}
  \centering
  \begin{minipage}{0.48\linewidth}
    \includegraphics[width=\textwidth]{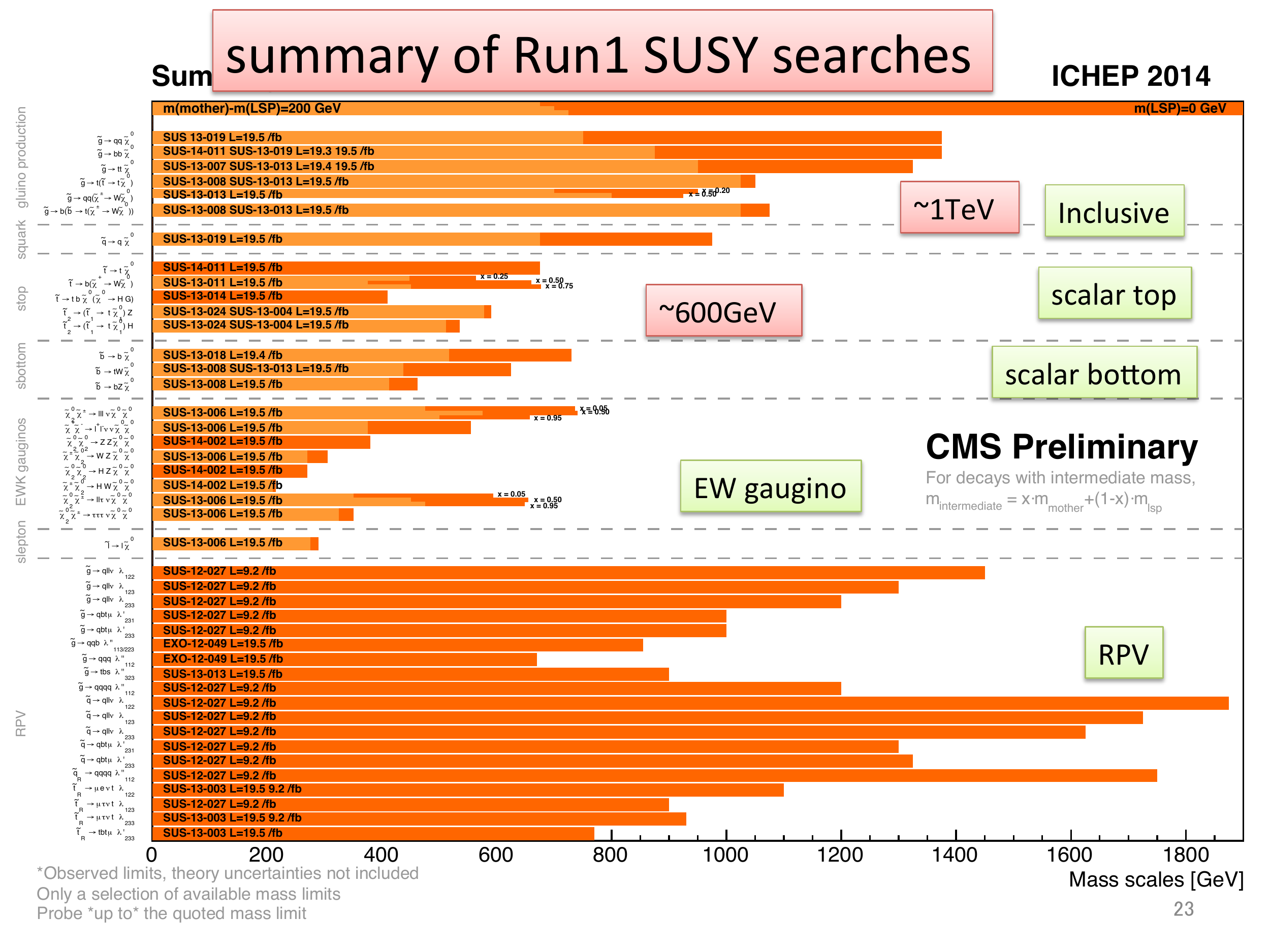}
  \end{minipage}\hfill
  \begin{minipage}{0.48\linewidth}
    \includegraphics[width=\textwidth]{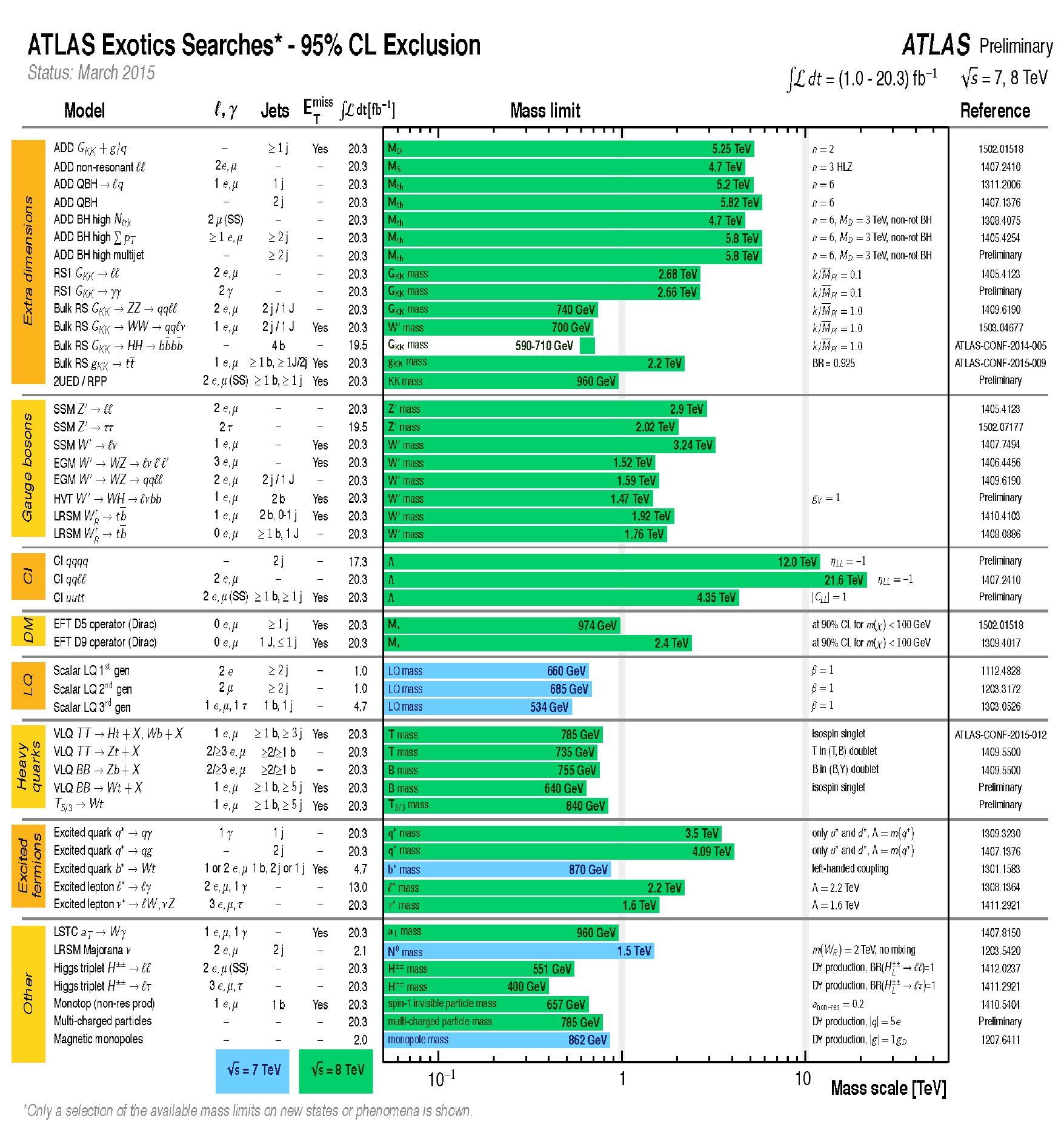}
  \end{minipage}
  \caption{Limits on new particle masses in SUSY searches (taken from talk by
    Asai\,\protect\cite{AsaiProceedings}) and
    ``exotics'' searches (taken from talk by
    Alcaraz\,\protect\cite{AlcarazProceedings}).
  }
  \label{fig:search-results}
\end{figure}

That headline figure perhaps obscures the fact that SUSY is not a
theory with a uniquely predicted spectrum. 
There are many SUSY partners to search for and, depending on the way in
which supersymmetry is broken, a range of possible mass spectra.
As a result, experimentally, SUSY searches get broken up into very
many channels.~\cite{AsaiProceedings} 
While some ``headline''
limits, notably on (degenerate) squarks and gluinos, approach 1.5\,TeV, the limit
on stop squarks is instead in the range $600-700\GeV$, while those
for electroweak SUSY partners can be even lower, cf.\
Fig.~\ref{fig:search-results} (left).

Searches for scenarios other than SUSY generally get classified as
``exotic'' searches.
This encompasses a range of new particles and phenomena such as heavy
gauge bosons (not so exotic!), leptoquarks, excited fermions, large
extra dimensions, RS gravitons, compositeness, various dark matter
candidates, etc., as illustrated in Fig.~\ref{fig:search-results}
(right).\,\cite{AlcarazProceedings}
One class of searches that I will highlight is that for displaced
jets, produced by decays of (relatively) long-lived new particles.
Between them, ATLAS and CMS and have managed to look jets originating
anywhere between a few millimetres from the collision point, all the
way up to several metres. 
Such searches were certainly not the main focus of the original design of
the LHC experiments, and the fact that they have been carried out so
successfully is a tribute to the ingenuity of the experiments in
triggering and exploiting their detectors.

\begin{figure}
  \centering
  \includegraphics[width=0.5\textwidth]{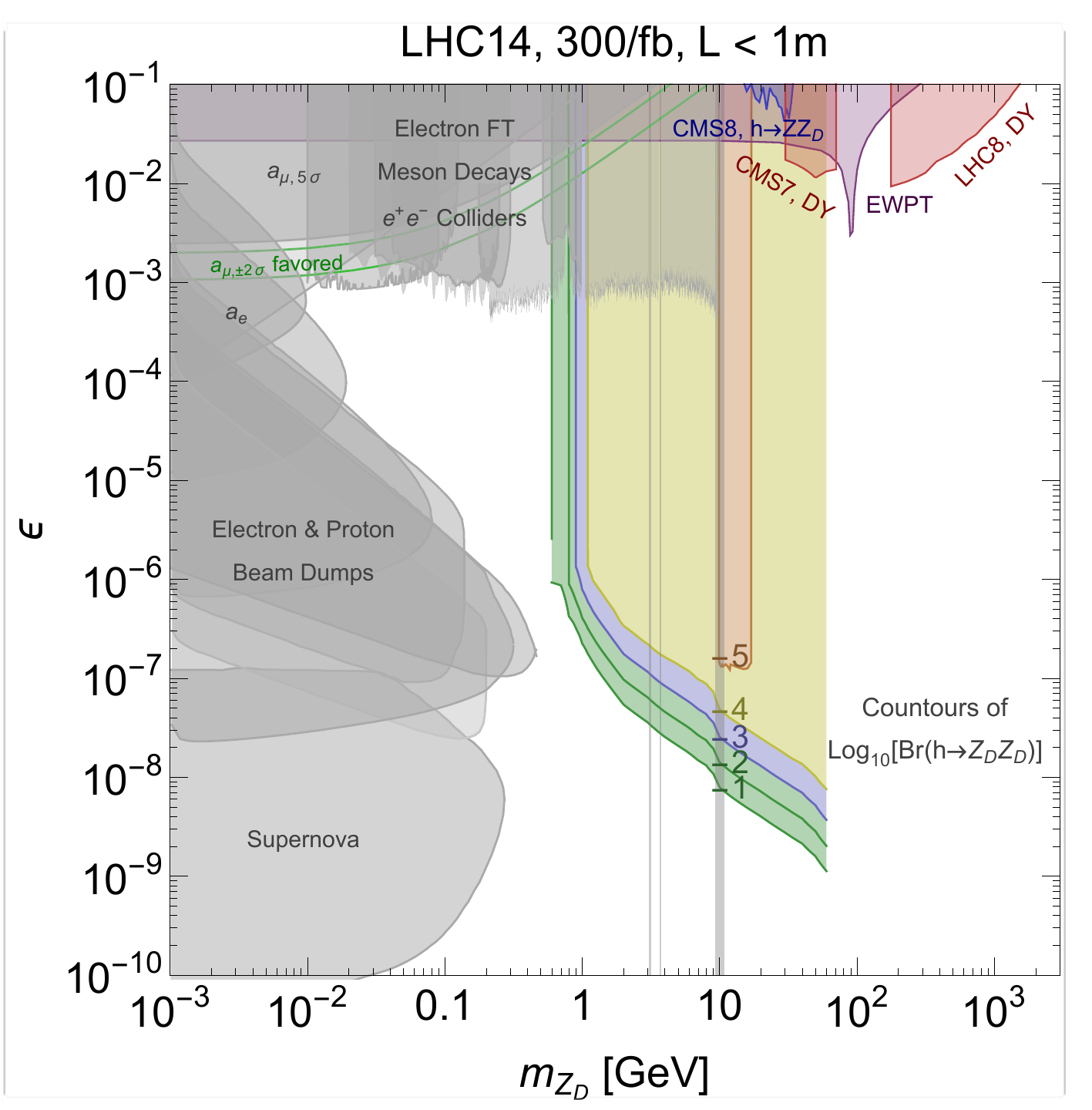}
  \caption{The LHC's potential for searches for $Z_D$ dark bosons using
    (possibly displaced) decays of Higgs bosons (figure taken from
    the talk by Shelton\,\protect\cite{SheltonProceedings}).}
  \label{fig:two-dark-Z}
\end{figure}

Another dimension that is opening up for searches involves the use of
the Higgs boson.
As emphasised by Shelton,\,\cite{SheltonProceedings} one of the respects
in which the Higgs is special is its very narrow width, about
$4.1\MeV$. 
This is to be compared to the $1-2\GeV$ width of all other
electroweak-scale particles.
A consequence of the narrow width is that new light degrees of freedom
with even only a tiny coupling to the Higgs can still be present in
appreciable fractions among its decays.
One example involved a Higgs decaying to two dark Z bosons ($Z_D$),
which mix with the electroweak sector to then decay to
leptons. 
The potential coverage from LHC searches is huge and very
complementary to other search approaches, as illustrated in
Fig.~\ref{fig:two-dark-Z}.

While the LHC experiments have searched for very many scenarios of new
physics, it is impossible for them to cover all possible theoretical
models, let alone models that are proposed after a given search is
complete. 
In contrast with astronomy, cosmic-ray and cosmology experiments, the
LHC data are generally not public: the largest release of data comes
from CMS, and involves about 0.2\% of their
dataset.\cite{CMS-data-release} 
There are various reasons for this, connected both with the traditions
of high-energy physics and the considerable complexity of the raw
data: a correct analysis even of the $0.2\%$ of public CMS data is,
for an outsider, almost certainly not a trivial enterprise.
Consequently, when someone proposes a new model, they cannot simply
compare it to data to see if it has already been excluded.
Instead it has become standard to adopt a ``recasting''
procedure:\,\cite{PapucciProceedings} for a given new model A, one
identifies existing LHC searches for another model (say B) with
similar signatures; one then generates Monte Carlo simulated events
for model A, and applies the same cuts that had been used to search
for B (including detector smearing and inefficiencies) and sees how many
events from model A would survive those cuts.
If that number is larger than the upper limit on the number of allowed
events in model B, then one can deduce that model A has been
excluded. 
Recasting appears to be a very powerful way to preserve the legacy of
LHC's searches.\footnote{Though there will always be some questions
  that can only be answered by a full reanalysis of the data.} 
Currently it is mostly being carried out by small groups of theorists,
with a few attempts ongoing to systematically recast a large number of
the LHC results.
In the long term, however, it is unclear whether any single small
group can keep pace with the many searches coming from the LHC over
its lifetime.
How best to scale and sustain the effort needed for generalised
recasting remains an open question for the field.



\begin{figure}
  \centering
  \includegraphics[width=0.45\textwidth]{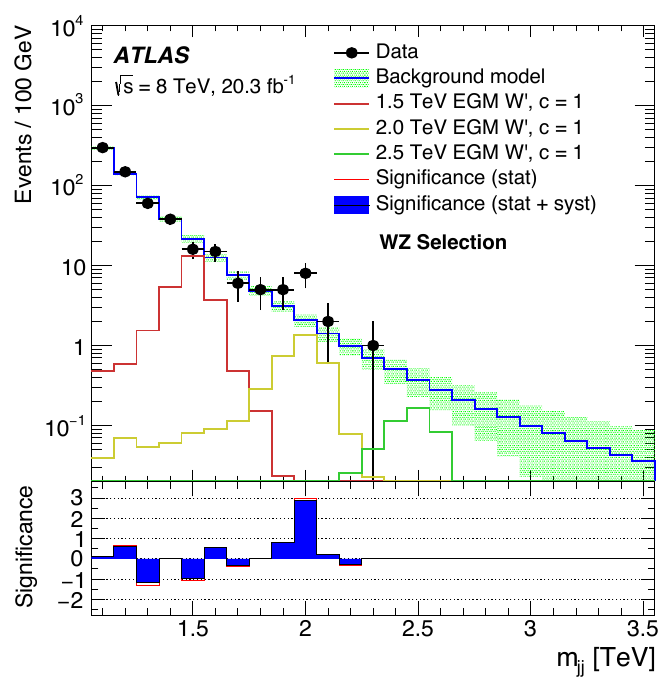}\hfill
  \includegraphics[width=0.48\textwidth]{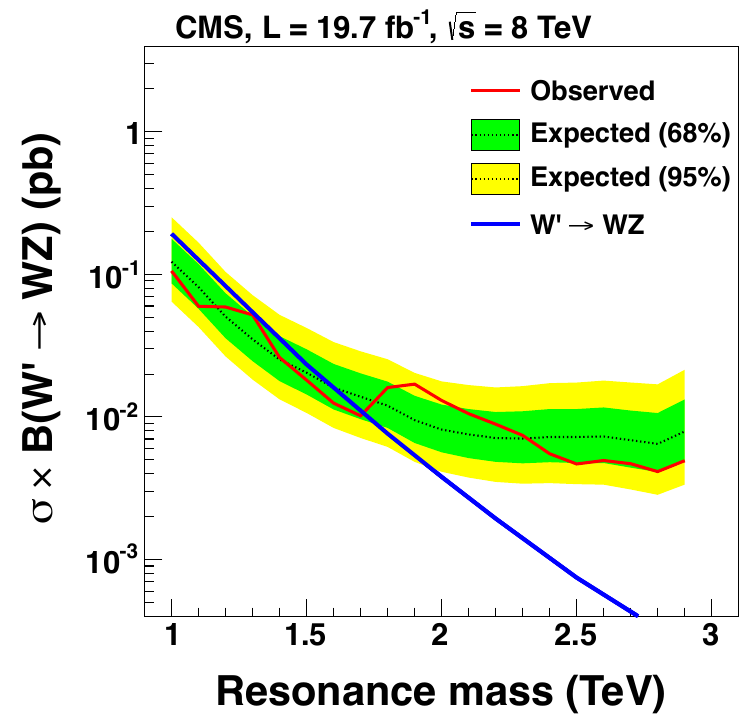}
  \caption{Results on the searches for diboson resonances in fully
    hadronic decay channels. Left, the
    number of events observed by ATLAS as a function of the dijet
    mass, with two W/Z-boson tags,\,\protect\cite{Aad:2015owa}
    displaying a prominent 
    bump around $2\TeV$.
    Right, the upper limit on the cross section $W' \to WZ$ from
    CMS,\protect\cite{Khachatryan:2014hpa} also showing an excess
    close to $2\TeV$.  }
  \label{fig:diboson}
\end{figure}

It would be impossible to close the section on searches without asking
whether there are any hints of new physics lurking in the existing
data.
The general answer is that while there are some discrepancies, there
is nothing compelling.
One example is in a channel with missing transverse momentum, jets and
two leptons (consistent with the $Z$ mass), where ATLAS observes a
$3\sigma$ excess in the electron channel and $1.7\sigma$ in the muon
channel. 
CMS however does not see an excess in the same place.
Another example (which has generated considerable theoretical
speculation since it appeared) is the search for a resonance decaying
to two vector bosons, where both ATLAS and CMS see hints of bumps
around 2 TeV, Fig.~\ref{fig:diboson}.
In the case of the ATLAS data, in the $WZ$ tagged channel, the excess
is $3.4\sigma$ locally and $2.5\sigma$ globally.
The CMS bump is slightly lower in mass than the ATLAS one, with a
somewhat lower cross section and significance.
%


\section{LHC Run 2 (and beyond)}

As the conference was proceeding, the LHC was gearing up for the
physics of Run 2, and first $13\TeV$ collisions with stable beams took
place midway through the week.
Lamont discussed the huge consolidation effort that went into
preparing the LHC for 13\TeV\ collisions.\cite{LamontProceedings}
He also described some of the challenges that they have encountered
during startup, including the retraining of the superconducting
magnets, so called ``UFOs'', and so forth. 
One message that emerged from his talk was ``this is not bad'': in a
project of this magnitude it is normal to encounter some difficulties
and so far those difficulties are being successfully handled as they
arise.
Nevertheless, he cautioned that with $6.5\TeV$ beams, the machine will
be operating much closer to its limits than was the case for $4\TeV$
beams.

In terms of luminosities, the hope for this year is to obtain between
about $4\fb^{-1}$ (recall Run 1 delivered $20\fb^{-1}$ at $8\TeV$)
and $100\fb^{-1}$ by the end of Run~2.
Alcaraz\,\cite{AlcarazProceedings} used a tool called
ColliderReach\,\cite{ColliderReach} to help illustrate when $13\TeV$
data will start to become competitive with $8\TeV$ results.
In searches involving large system masses, say around $4\TeV$ (e.g. the
current limit for excited quarks), just $0.1\fb^{-1}$ will be
sufficient.
At lower masses, say $1\TeV$, one needs about $5\fb^{-1}$ to overtake
the Run~1 results.

\begin{figure}
  \centering
  \includegraphics[width=0.48\textwidth]{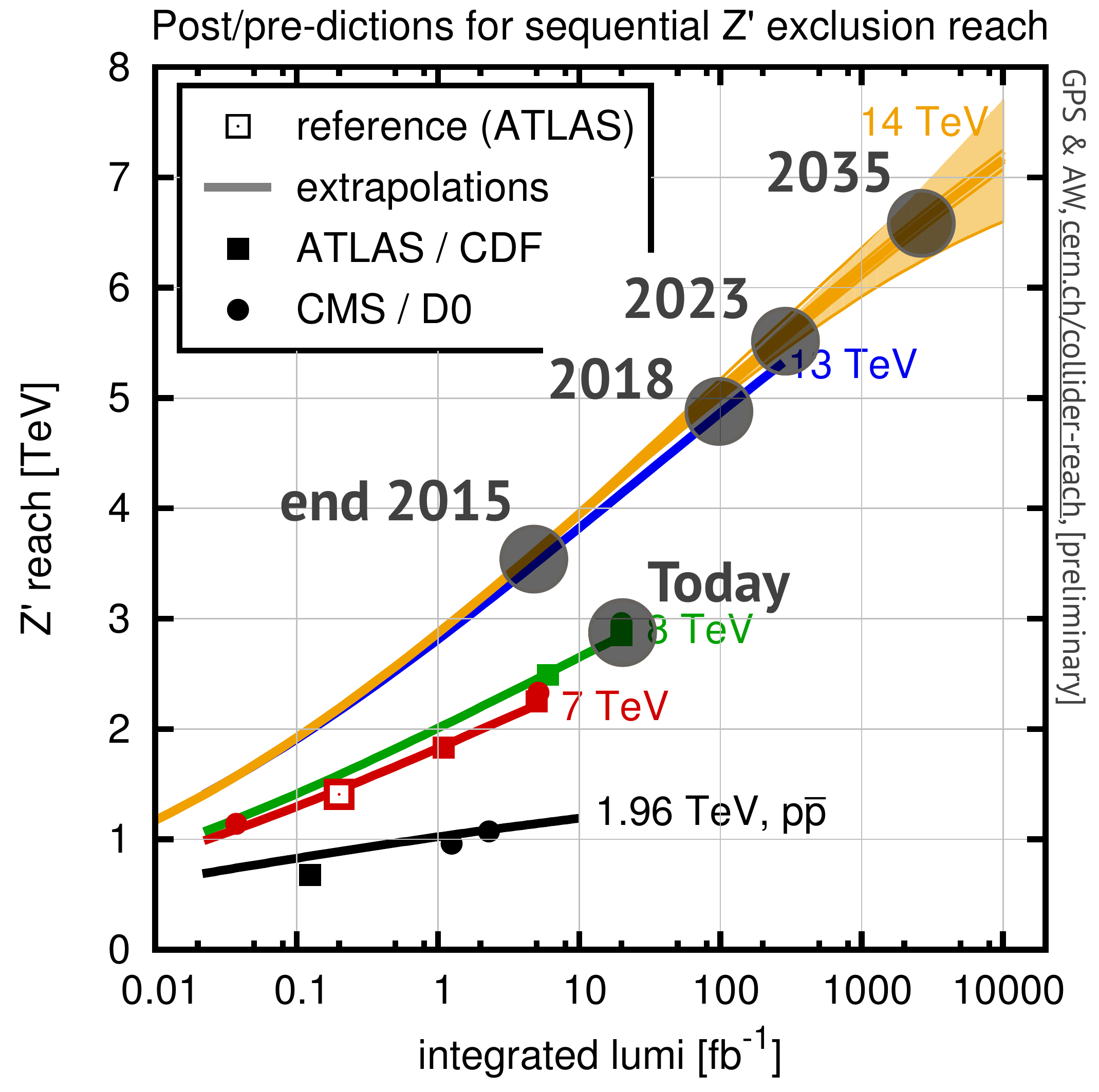}
  \caption{Post- and predictions for the $Z'$ search (exclusion)
    sensitivity, as a function of luminosity, for different collider
    setups and centre-of-mass energies.\,\protect\cite{ColliderReach}
    The (post)prediction is based on the momentum dependence of
    partonic luminosities.
    Key luminosity targets for the LHC programme are highlighted
    together with approximate dates when they should be achieved.
  }
  \label{fig:zprime-reach}
\end{figure}

To look further into the future, it is instructive to take a simple,
concrete example, say a sequential standard model $Z'$ decaying to
leptons (Fig.~\ref{fig:zprime-reach}). Today the limit is about
$2.9\TeV$. Using the ColliderReach tool for extrapolations, one can
establish that by the end of 2015 with $5\fb^{-1}$ the limit could go
up to $3.6\TeV$;
by the end of Run~2 in 2018, with $100\invfb$, this should rise to
about $5\TeV$;
Run~3 ($300\invfb$, 2023) should take this to $5.4\TeV$, while the
high-luminosity LHC ($3000\invfb$, around 10 years later), will take
us to about $6.4\TeV$.
The details vary depending on the precise search, but a common pattern
that emerges is that the coming year offers only the very first step
towards the ultimate limit of what LHC will be able to probe, even if
some patience may be needed on the way towards that
limit. 

\section{(Quark) Flavour Physics}

\begin{figure}
  \centering
  \begin{minipage}{0.48\linewidth}
    \includegraphics[width=\textwidth]{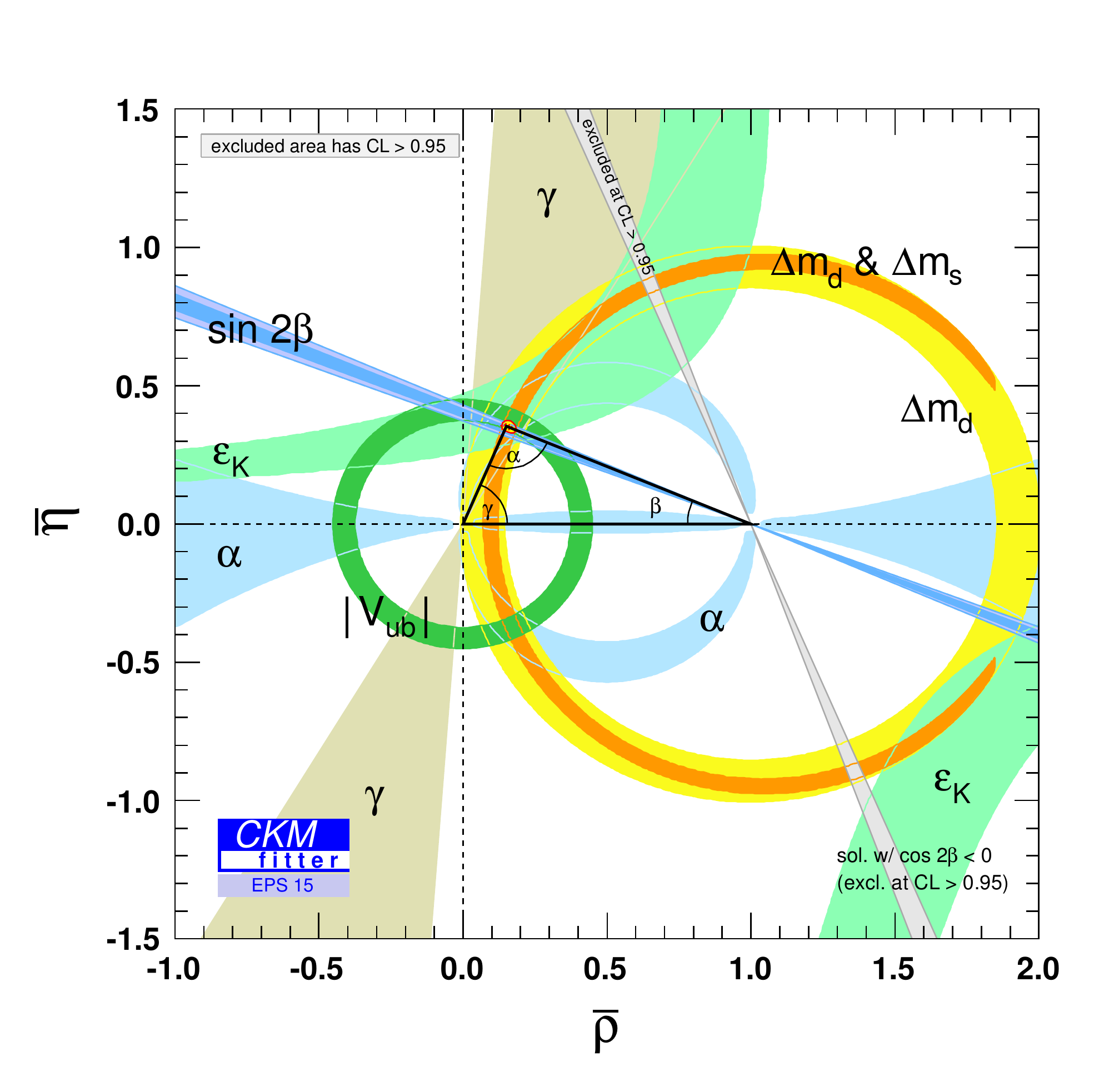}
  \end{minipage}\hfill
  \begin{minipage}{0.48\linewidth}
    \includegraphics[width=\textwidth]{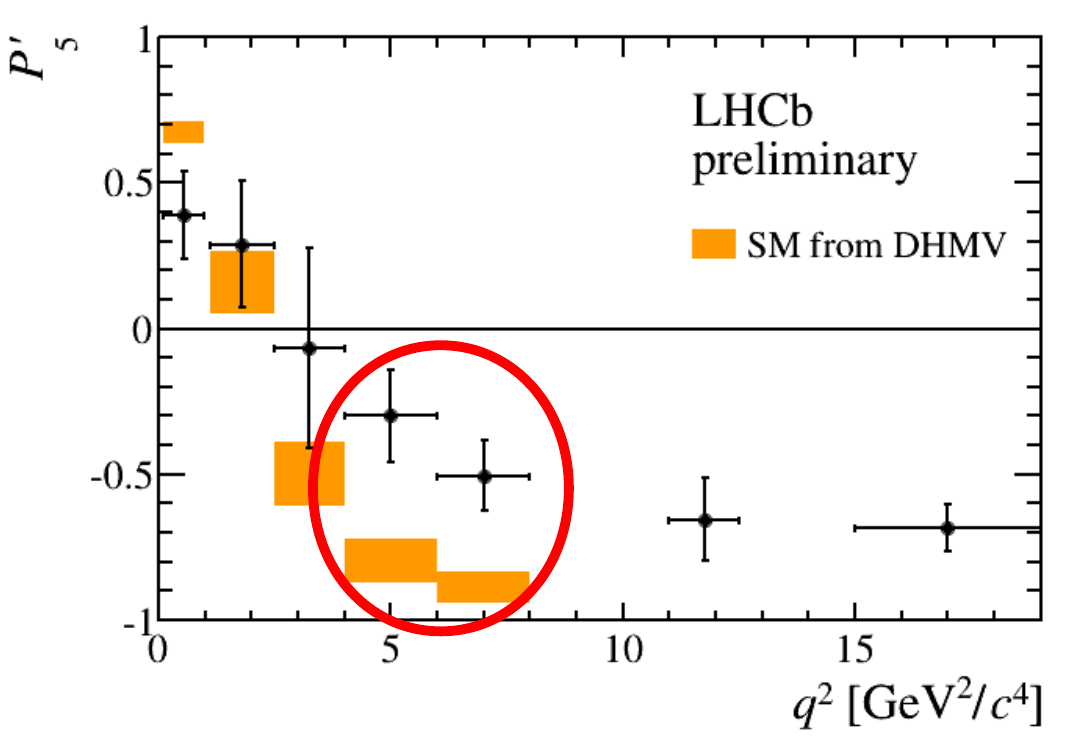}
  \end{minipage}
  \caption{Left: status of constraints on the CKM matrix (plot as shown in
    the talk by Gershon,\,\protect\cite{GershonProceedings} taken from
    Ref.~\protect\cite{CKMFitter}).
    Right: the distribution of the $P_5'$ angular observable in $B^0 \to
    K^{\star0}\mu^+\mu^-$ decays, as a function of the momentum scale
    $q^2$, highlighting the discrepancy around $5\GeV^2$ (figure taken
    from talk by Gershon\,\protect\cite{GershonProceedings}).
}
  \label{fig:ckm-constraints}
\end{figure}

So far, most of the discussion has concentrated on direct probes of
scales from a hundred GeV to a few TeV.
Yet almost $50\%$ of the conference was dedicated to subjects outside
this direct range of scales.
Experimentally, quark flavour physics mostly involves scales of a few
GeV. 
A huge effort from the flavour experiments (and also the lattice QCD
community) has led to multiply constrained determinations of the
different elements of the CKM matrix, with a generally consistent
picture emerging from many different measurements, cf.\
Fig.~\ref{fig:ckm-constraints} (left).
Some points of tension do persist, and in his review talk
Gershon~\cite{GershonProceedings} highlighted the difference between
$V_{ub}$ extractions from inclusive $B$-meson decays and exclusive
decays.
A new addition to this story was recent LHCb data on $V_{ub}$ from
exclusive decays of $B$-baryons, which is very consistent with that
from the exclusive meson decays.
Another place of tension is in lepton universality with the observed
ratio of $(B \to D^{(*)}\tau\nu)/(B \to D^{(*)}\mu\nu)$ decays being
$3-4\sigma$ higher than expected in the standard model, a finding that
has been reinforced by recent LHCb data.\,\cite{LHCb-PAPER-2015-025}

As well as determining the CKM matrix (and verifying the consistency
of different determinations) an important aspect of flavour physics
is the study of rare decays.
The $B_s \to \mu^+\mu^-$ is one decay that that is of particular
interest because it is highly suppressed in the standard model, while
it may be enhanced in new-physics models.
This is similar to the argument, given above, about Higgs decays being especially
interesting places to search for new physics, because of the narrow
width of the Higgs boson.
The $B_s \to \mu^+\mu^-$ branching ratio has recently been measured,
jointly by CMS and LHCb,\,\cite{NatureBsmumu} to be about $3\times
10^{-9} \pm 20\%$, consistent with the standard model.
Haisch, in his review,\,\cite{HaischProceedings} made several
comparisons with Higgs physics: the experimental precision
$B_s \to \mu^+\mu^-$ branching ratio is comparable to that on Higgs
production and decay; what's more if one tries to deduce a limit on
the scale of new physics based on the observed consistency with the
standard model, one reaches conclusions that are similar, just below a
TeV (in the case of $B_s \to \mu^+\mu^-$ the limit becomes much higher if one
relaxes the assumption that new physics has a flavour structure aligned
with that of the standard model, i.e.\ minimal flavour violation).

One point of tension in rare decays that is currently the subject of
much attention is the so-called $P_5'$ angular observable in
$B^0 \to K^{*0} \mu^+\mu^-$
decays,\cite{GershonProceedings,LHCb-CONF-2015-002}
Fig.~\ref{fig:ckm-constraints} (right).
There was some discussion\,\cite{HaischProceedings} however as
to the degree of robustness of the theoretical predictions for this
observable.

The difficulties that arise in interpretations of hadronic physics
were discussed also in the context of studies of hadronic resonances that
are candidates for being four-quark bound states.\cite{OlsenProceedings}
In the future it may be possible to obtain complementary probes of
anomalies in the quark sector using high-momentum-transfer processes,
and one in particular that was highlighted was $t\bar t Z$
production.\cite{SchulzeProceedings}

\section{Neutrinos and the lepton sector}

\begin{figure}
  \centering
  \includegraphics[width=0.55\textwidth]{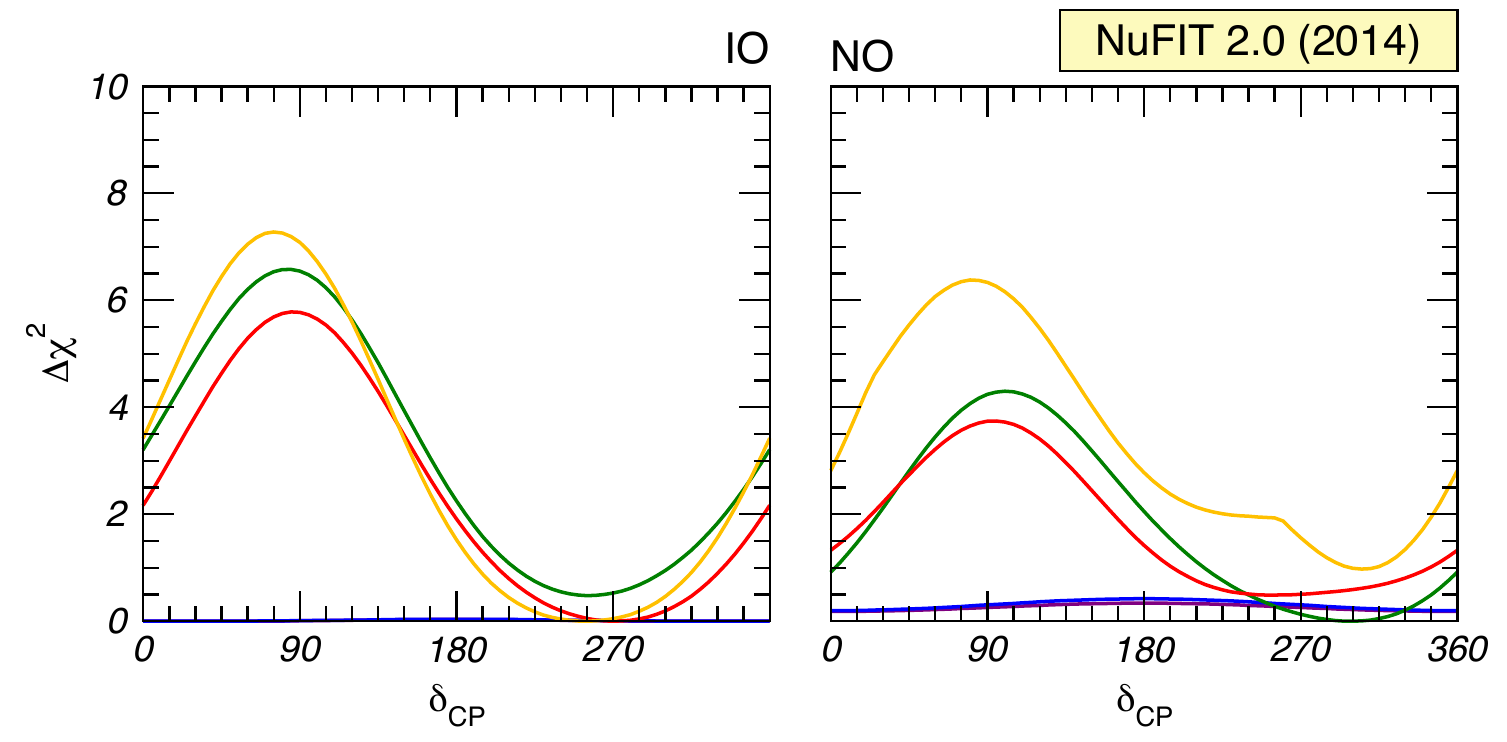}
  \caption{NuFIT results showing a $2\sigma$ hint for a CP phase
    around $\delta_\text{CP} = \frac{3\pi}{2}$. Taken from the talk by
  Gonzalez-Garcia.\protect\cite{GonzalezGarciaProceedings}}
  \label{fig:nufit-CP-violation}
\end{figure}

The neutrino sector of particle physics is special in that a number of
the important unanswered questions should be clearly resolvable in the
coming decade.
The state of today's
knowledge~\cite{GonzalezGarciaProceedings,ZitoProceedings} is that we
have reasonable constraints on the absolute values of the neutrino
mixing matrix $V_\text{PMNS}$
\begin{equation}
  \label{eq:1}
  V_{\text{PMNS}} \simeq 
  \left( 
    \begin{array}{ccc}
      0.8 & 0.5 & 0.2\\
      0.4 & 0.6 & 0.7\\
      0.4 & 0.6 & 0.7
    \end{array}
  \right),
  \qquad
  V_{\text{CKM}} \simeq 
  \left( 
    \begin{array}{ccc}
      1  & 0.2 & 0.001\\
      0.2 & 1 & 0.01\\
      0.001 & 0.01 & 1
    \end{array}
  \right),
\end{equation}
here compared to the absolute values for the corresponding quark
mixing matrix $V_\text{CKM}$, and illustrating the much greater mixing
in the neutrino sector.
There are $2\sigma$ hints for a large CP-violating phase,
Fig.~\ref{fig:nufit-CP-violation}, whose value
could be of
importance for understanding the origin of the baryon asymmetry of the
universe.

Regarding neutrino masses, there are bounds from $\beta$-decay
experiments on the electron neutrino mass, $m_{\nu_e} < 2.2\eV$.
Bounds from cosmology depend on the specific cosmological observables
being considered, but can be as strong as
$\sum m_\nu \lesssim 0.17 \eV$ on the sum of neutrino masses.
Oscillations provide information on mass differences:
$\delta m^2_{12} \simeq 7\times 10^{-5} \text{eV}^2$ and
$|\delta m^2_{23}| \simeq 2\times 10^{-3}\,\text{eV}^2$.
The sign of $\delta m^2_{23}$ is however not known, so the problem of
determining the mass hierarchy remains open.
It is also not known whether neutrinos are Majorana or Dirac fermions.

\begin{figure}
  \centering
  \includegraphics[width=0.31\textwidth]{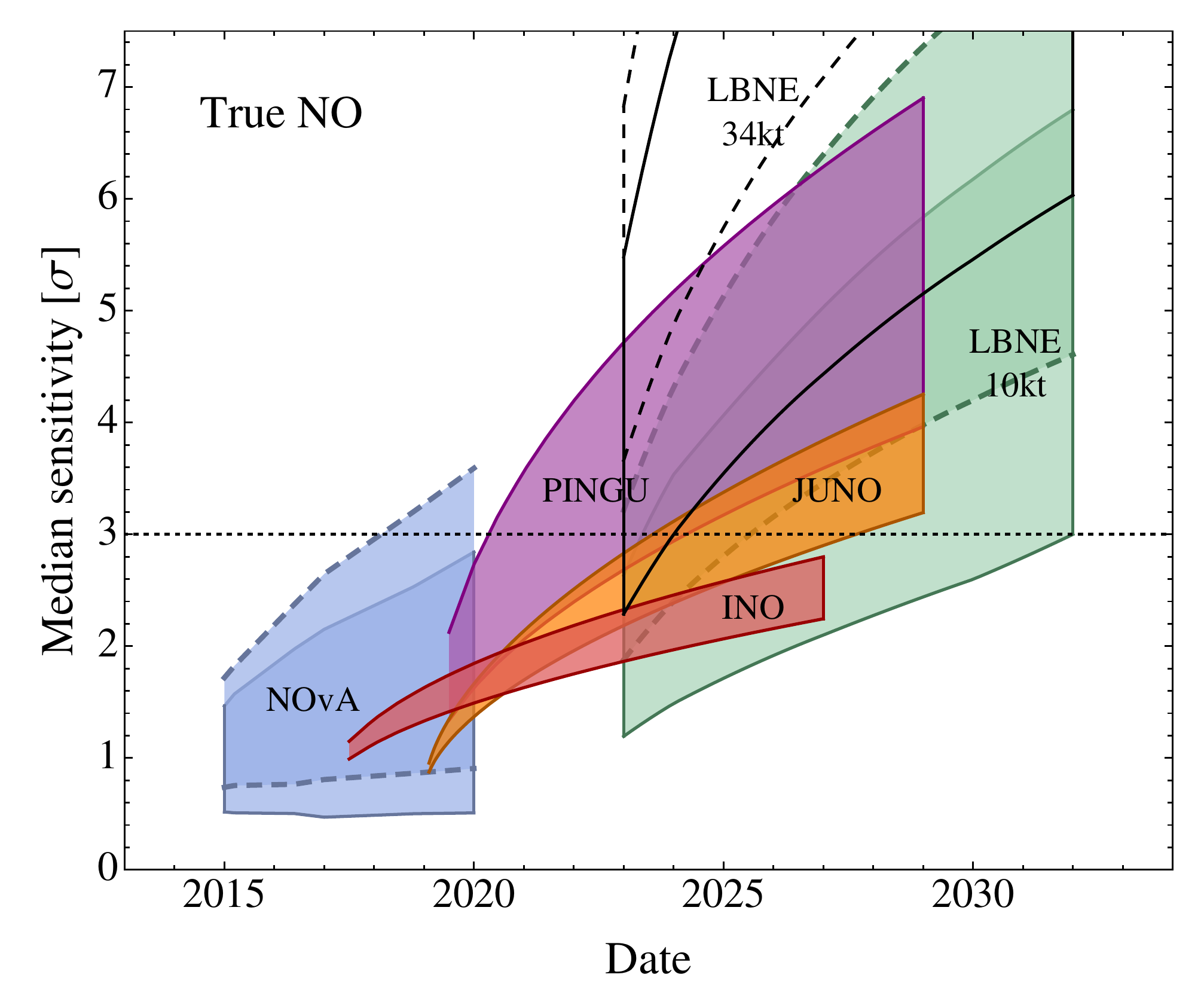}
  \;
  \includegraphics[width=0.31\textwidth]{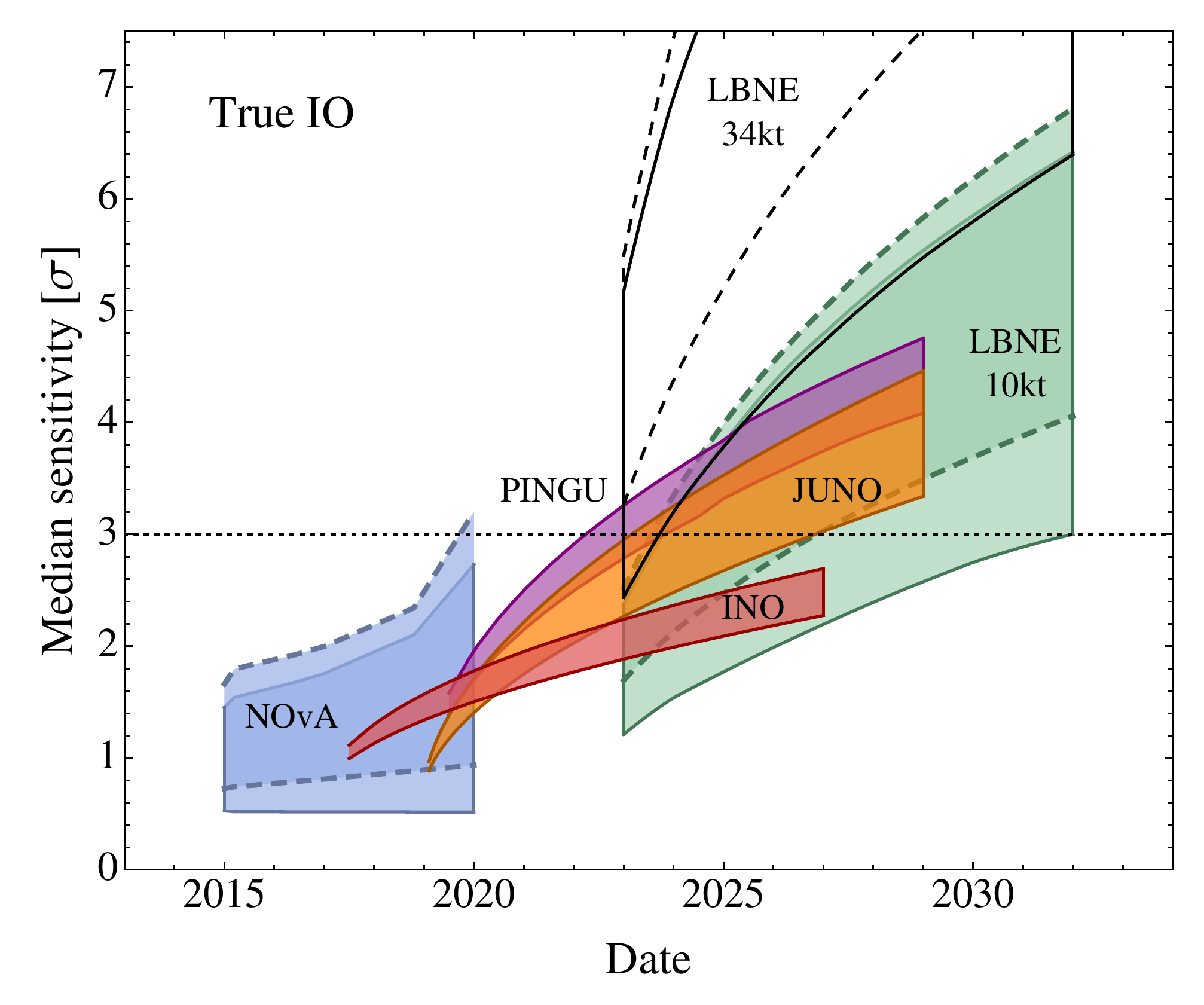}\;
  \includegraphics[width=0.31\textwidth]{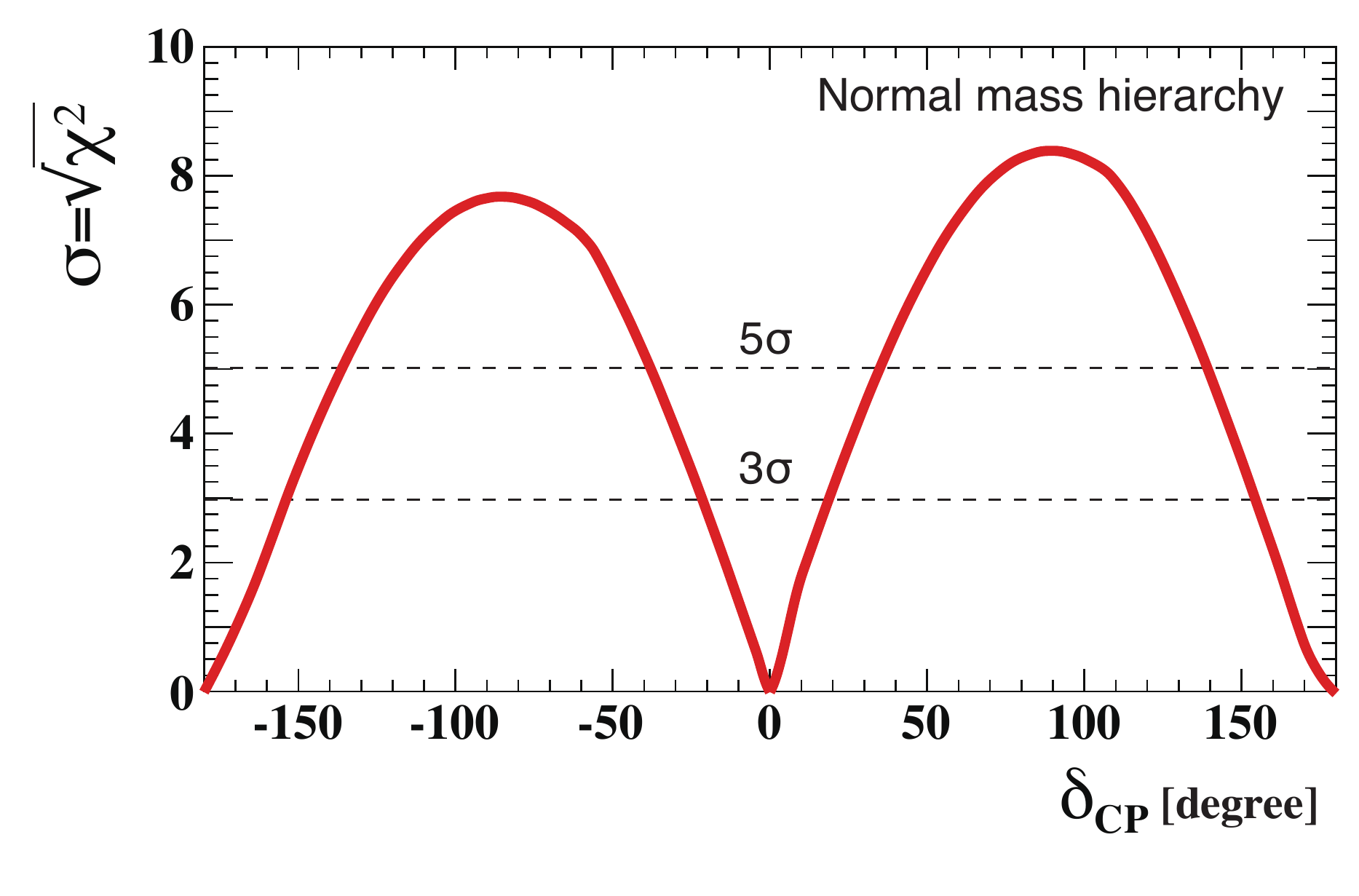}
  \caption{Left and middle: timeline for the expected range of
    significances for the determination of the mass ordering hierarchy
    of the neutrino sector, illustrating the different contributing
    experiments in both a normal (left) and inverted
    (middle) hierarchy.\protect\cite{Blennow:2013oma} 
    Right: expected significance on the CP-violating phase
    $\delta_\text{CP}$ from the Hyper-Kamiokande experiment, as a function
    of $\delta_\text{CP}$.\protect\cite{Abe:2015zbg} 
  }
  \label{fig:neutrino-MO-timeline}
\end{figure}

One new result presented at this conference concerned the observation
of $\bar \nu_\mu$ disappearance by T2K.\cite{QuilainProceedings}
The results are compatible with those for $\nu_\mu$ disappearance, as
should be the case assuming CPT symmetry.
To make progress with the determination of the CP-violating phase, it
is necessary to observe a difference in the rates of
$\nu_\mu \to \nu_e$ and $\bar \nu_\mu \to \bar \nu_e$ oscillations.
However neutrino-matter interactions also induce such an asymmetry,
and the magnitude of this effect depends on the (unknown) neutrino
mass hierarchy.
To disentangle the two effects requires an appropriate span of
neutrino energies and oscillation baseline lengths.
An indicative timeline for the evolution of different experiments'
sensitivities to the mass hierarchy is shown in
Fig.~\ref{fig:neutrino-MO-timeline},
as is the  sensitivity to the CP-violating phase that should eventually come,
on a 10-year timescale, from the Hyper-Kamiokande experiment (the LBNF/DUNE
experiment will also provide similar information). 

Regarding the absolute mass scale, progress is expected from the
Katrin experiment on the limit on the electron neutrino mass.
If neutrinos are Majorana fermions, then there is also a prospect of
sensitivity to this in neutrinoless double-$\beta$ decay experiments
in the next decade, notably for an inverted mass
hierarchy.\cite{TomeiProceedings,WalterProceedings}

Neutrino-based studies of the lepton sector are complemented by
experiments with charged leptons, where substantial improvements, for
example in limits on $\mu^+ \to e^+e^+e^-$ decays, are expected in the
coming years.\cite{FeruglioProceedings}

\section{Cosmic Rays}

\begin{figure}
  \centering
  \begin{minipage}{0.53\linewidth}
    \includegraphics[width=\textwidth]{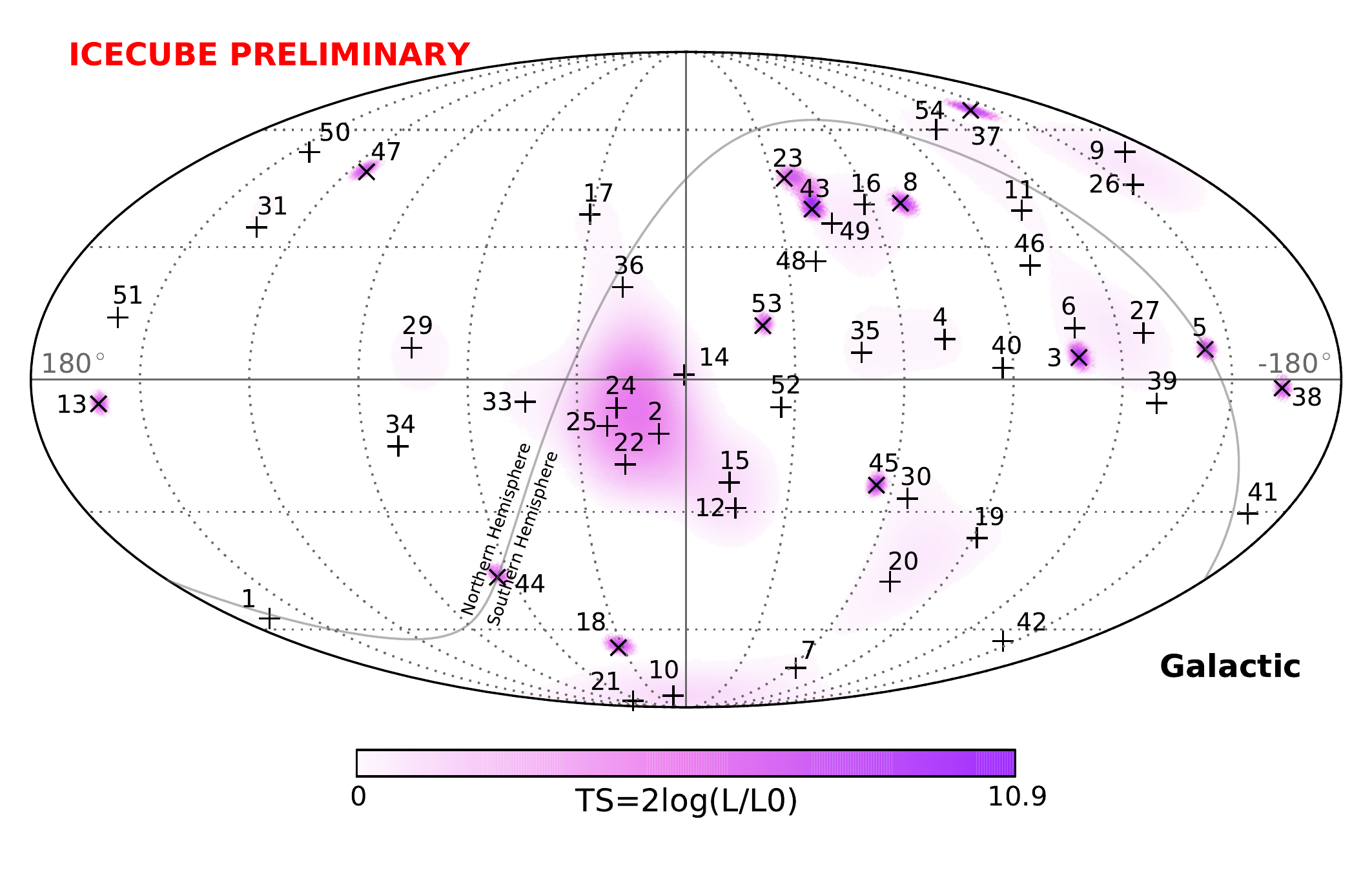}
  \end{minipage}\hfill
  \begin{minipage}{0.43\linewidth}
    \includegraphics[width=\textwidth]{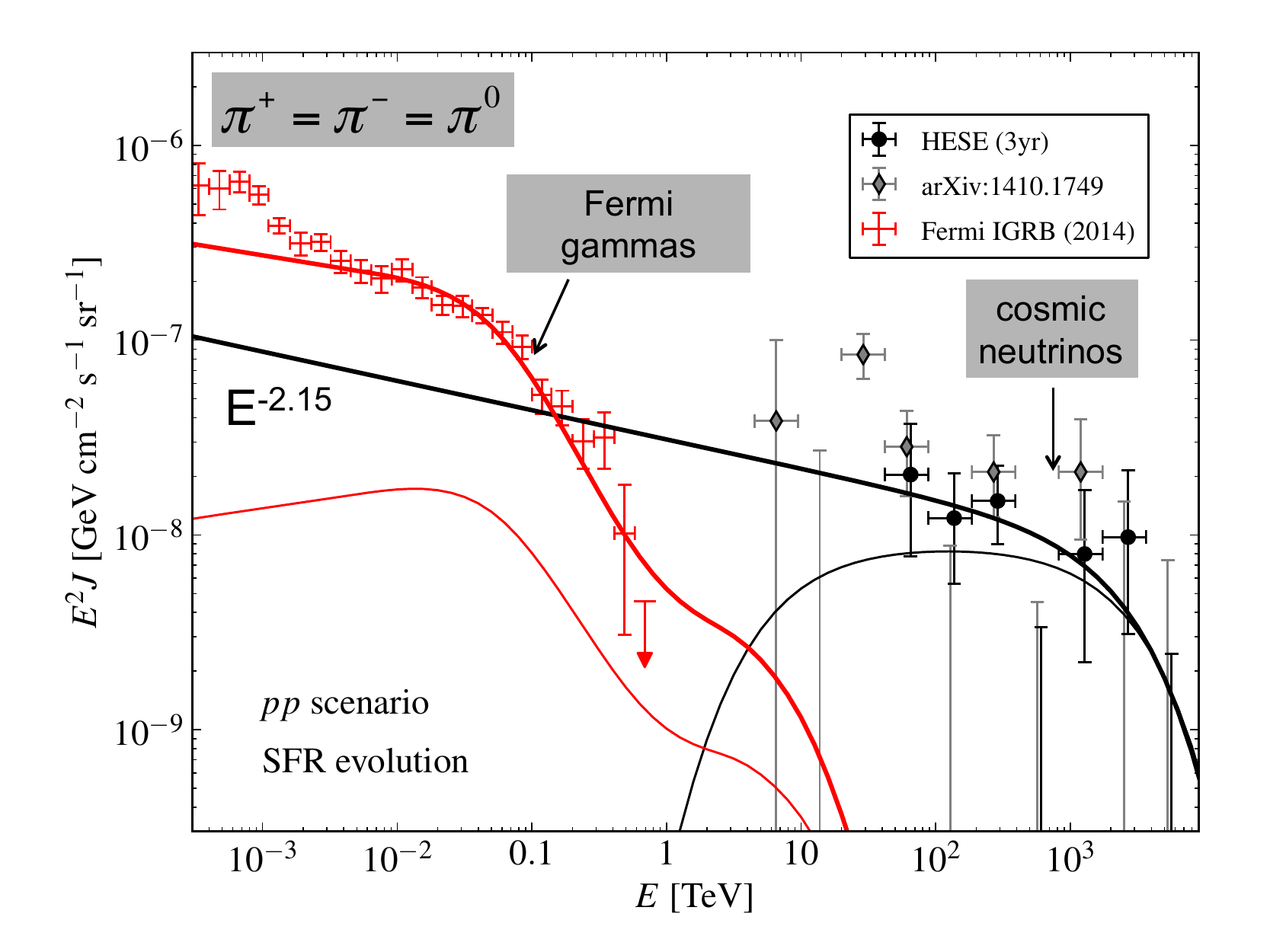}
  \end{minipage}
  \caption{
    Left: distribution of neutrinos from the sky as reported by
    the IceCube experiment; the shading represents a test statistic
    (TS) for clustering. 
    Right: spectra of photons (in red) and
    cosmic neutrinos (in black).
    The thick lines represent an attempt to jointly fit the neutrino
    and $\gamma$ ray spectra under the assumption that both arise
    dominantly from pion decay.
    Both plots taken from the talk by
    Halzen.\,\protect\cite{HalzenProceedings} }
  \label{fig:neutrino-astronomy}
\end{figure}

The subject of neutrinos naturally brings us to the question of cosmic
rays, where one of the significant advances in recent years has been the
observation of astrophysical neutrinos by the IceCube
detector, as presented by Halzen.\,\cite{HalzenProceedings}
The importance of neutrinos is that like photons, but unlike charged
cosmic rays, their direction of arrival points back to the source;
furthermore, in contrast to photons, they undergo essentially no
absorption or scattering as they travel to us.

The full set of astrophysical neutrino candidates is shown in
Fig.~\ref{fig:neutrino-astronomy} (left). 
Initial two-year data appeared to have an excess coming from the
galactic centre, however in the latest four-year results the
significance of that excess has gone down.
Currently there is no evidence for any other clustering within the
neutrino dataset.\footnote{Studies of correlations to catalogues of
  known sources is currently ongoing.}
To reliably observe multiple neutrinos from any single astrophysical
sources, it is expected that a $10\,\text{km}^3$ detector volume would
be sufficient.
Remarkably, the antarctic ice is sufficiently transparent that such a
detector could be successfully instrumented with the same number of
``strings'' as currently used for the $1\,\text{km}^3$ volume of
IceCube.

Halzen also showed a quantitative comparison of the observed neutrino
flux with the gamma-ray flux. 
The basis of the comparison is that a main expected source of
neutrinos is $\pi^\pm$ decays with the pions themselves being produced
for example in collisions of energetic protons with some kind of
target (e.g.\ gas).
In that case energetic photons should be produced at a similar rate in
$\pi^0$ decays.
However very energetic photons will then interact with the cosmic
microwave background (CMB), leading to a degradation of their energy.
The observed neutrino flux can therefore be used to infer a flux of
pion-decay induced gamma-rays, which, remarkably, coincides well
with the high-energy part of the observed Fermi gamma-ray spectrum,
Fig.~\ref{fig:neutrino-astronomy} (right).
This would suggest that it is production of pions in astrophysical
accelerators that is responsible for most of the photon flux.

\begin{figure}
  \centering
  \includegraphics[width=0.53\textwidth]{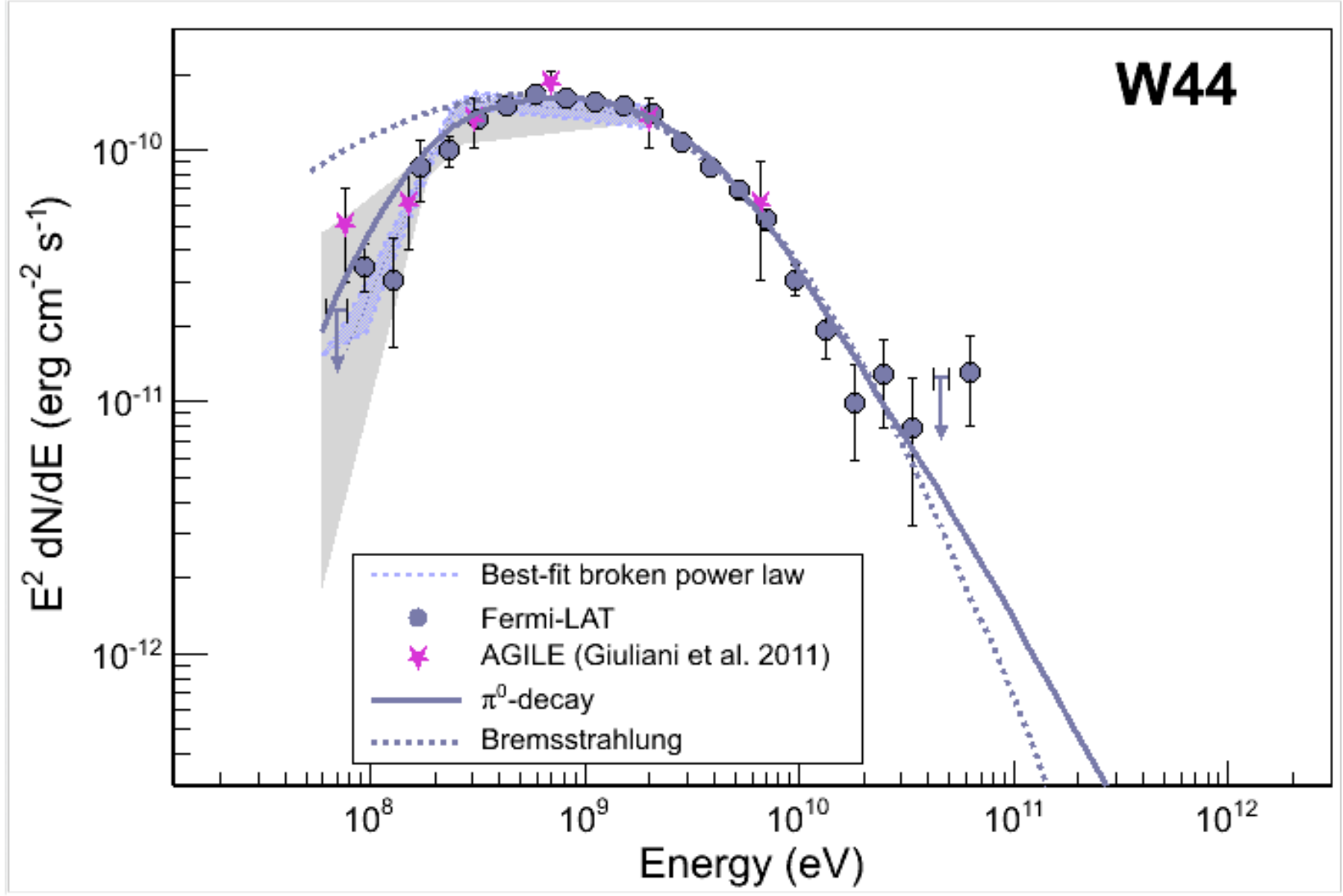}
  \caption{Photon spectrum from the W44 supernova remnant, compared to
    various models, including one in which the photons originate from
    $\pi^0$ decay. Plot taken from talk by Funk.\,\protect\cite{FunkProceedings} } 
  \label{fig:pi0-lower-threshold}
\end{figure}

The origin of energetic gamma rays was discussed also by
Funk.\,\cite{FunkProceedings}
He observed that if very high energy photons are being produced in
$\pi^0$ decays, then there should be a dip in the spectrum around
$m_\pi/2$.
Fig.~\ref{fig:pi0-lower-threshold} shows that such a dip does indeed seem to
be present in data from Fermi-LAT.

\begin{figure}
  \centering
  \begin{minipage}{0.53\linewidth}
    \includegraphics[width=\textwidth]{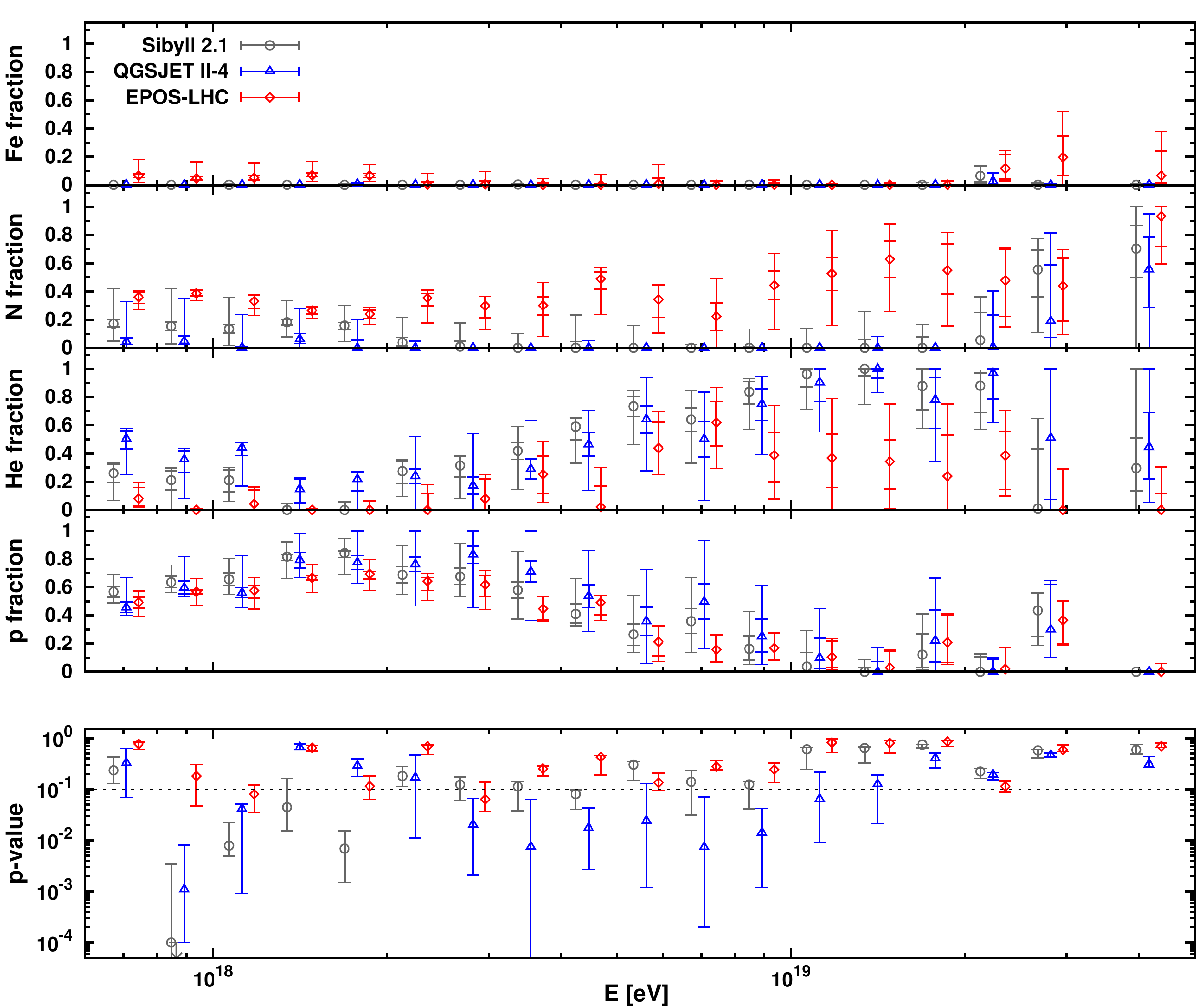}
  \end{minipage}\hfill
  \begin{minipage}{0.43\linewidth}
    \includegraphics[width=\textwidth]{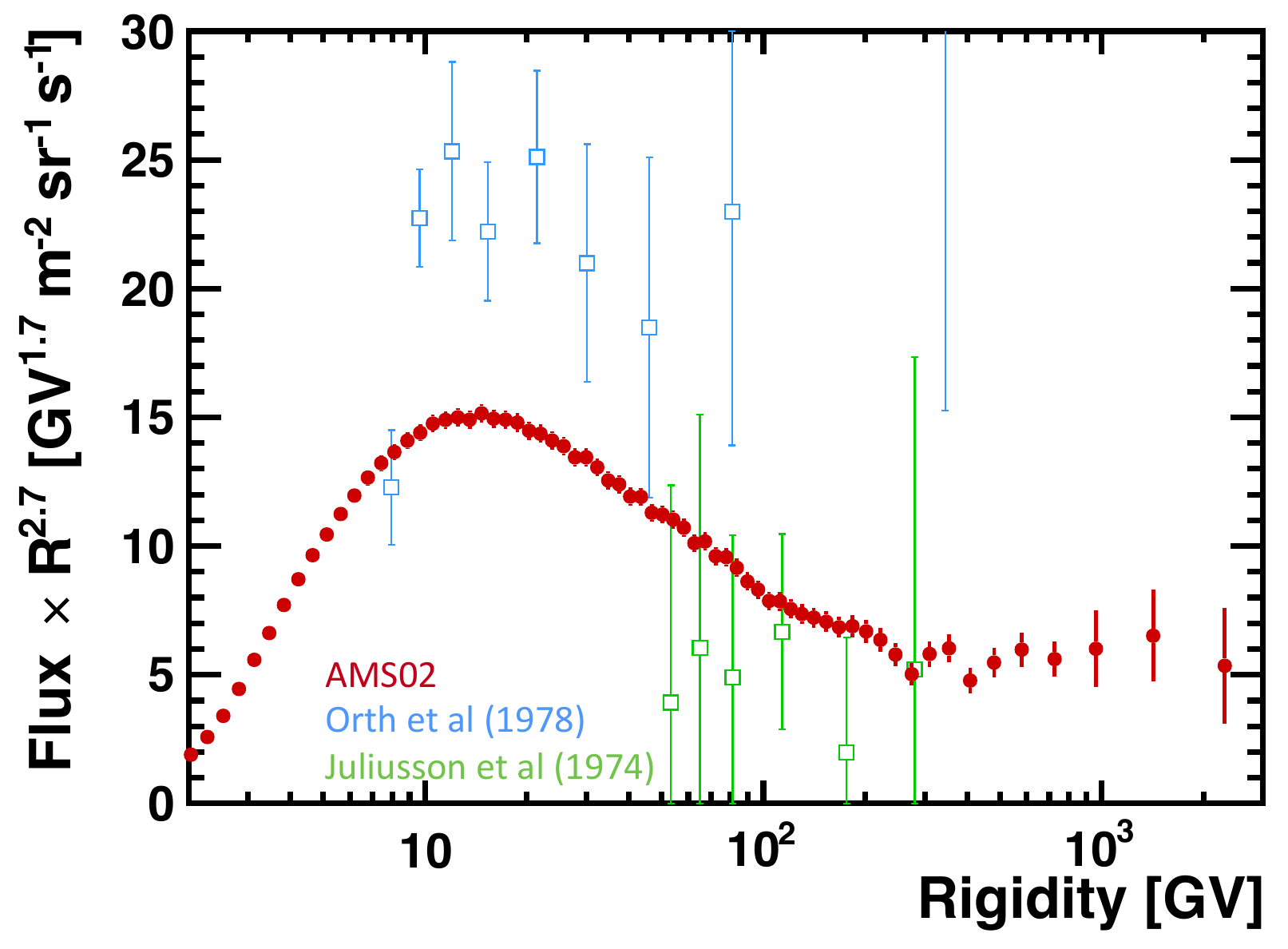}
  \end{minipage}  
  \caption{Left: fits for the nuclear composition of cosmic rays, based on the
    depth of the maximum of the cosmic ray shower, shown as a function
    of different energies, with various Monte Carlo programs used to
    model the shower development; taken from talk by Roth.\,\protect\cite{RothProceedings}
    Right: cosmic Lithium flux, shown as a function of  ``rigidity''
    ($p/Z$ where $p$ is the momentum) as measured by the AMS
    collaboration; taken from the talk by Derome.\,\protect\cite{DeromeProceedings}
  }
  \label{fig:CR-composition}
\end{figure}

Despite these hints about the origin of high-energy photons, much
remains to be understood as to how, precisely, cosmic rays are
accelerated to the very high energies that are observed. 
Most probably, this occurs in shocks in supernova remnants, but there
is as yet no observational proof.\,\cite{MarcowithProceedings}
Still, even if such proof is lacking, there is considerable progress
in learning at least about the composition of cosmic rays.
At the very highest energies, up to $\sim 10^{19}\eV$,
Roth\,\cite{RothProceedings} presented results from Auger on the
fraction of protons, helium, nitrogen and iron,
Fig.~\ref{fig:CR-composition} (left). 
At lower energies, up to about $1\TeV$, we saw beautifully precise
data on the fluxes of various nuclei as measured by the AMS
experiment,\cite{DeromeProceedings} e.g.\ Fig.~\ref{fig:CR-composition} (right).
%

\section{Dark Matter}

As well as being of intrinsic interest as a window on the
astrophysical mechanisms at play in the universe, cosmic rays offer
the prospect of indirect detection of dark matter.
Given some dark matter particle $\chi$, one can for example imagine
annihilations $\chi\chi \to W^+W^-, ZZ, q\bar q,$ etc., all of which
would result in a continuum distribution of protons, photons and
electrons that turns off somewhere below $m_\chi$.
There is also the possibility of annihilation such as $\chi\chi \to
\gamma \gamma$, which would give a distinct line signal in the
$\gamma$-ray spectrum.

\begin{figure}
  \centering
  \includegraphics[width=0.48\textwidth]{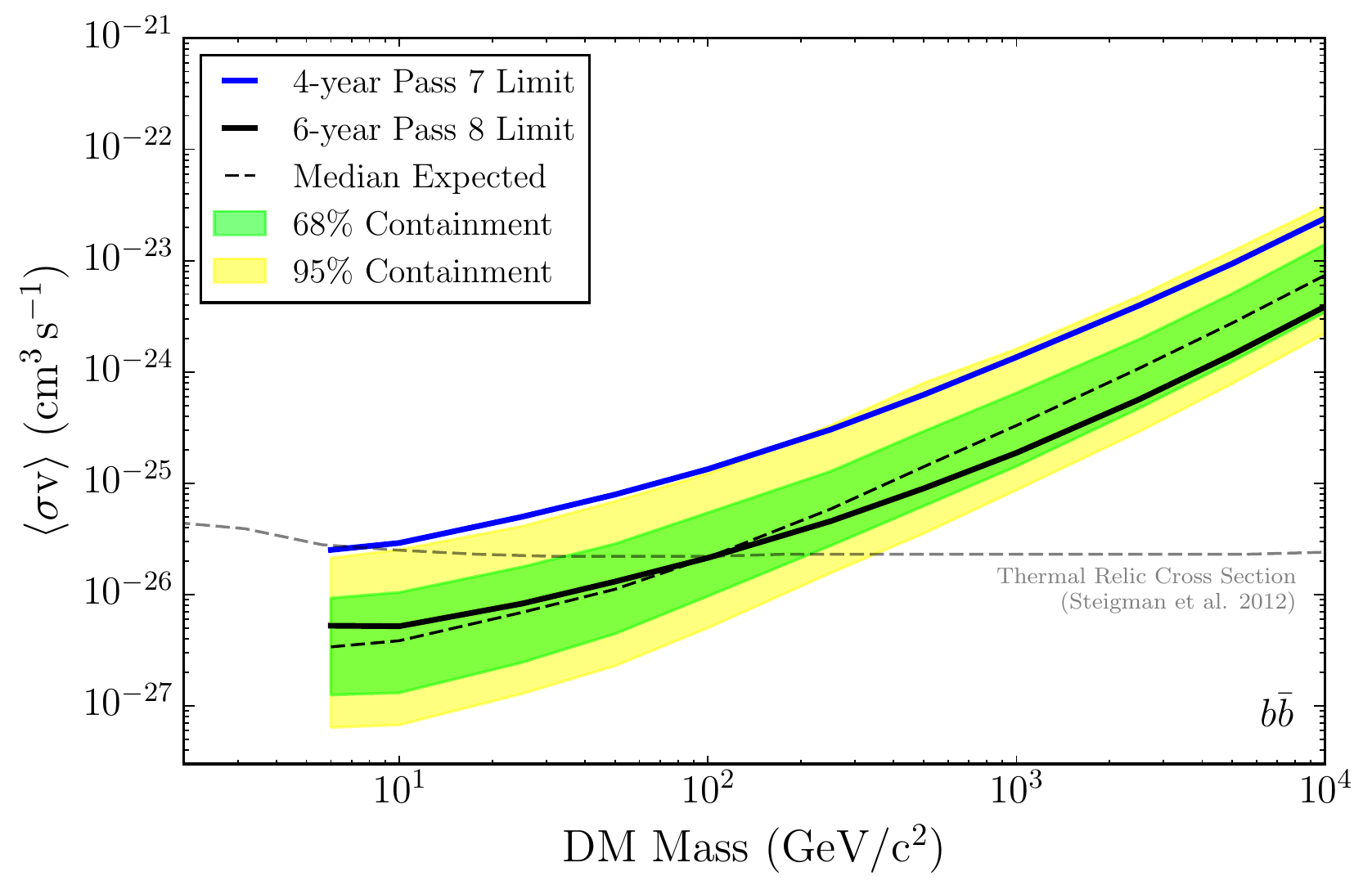}
  \hfill
  \includegraphics[width=0.48\textwidth]{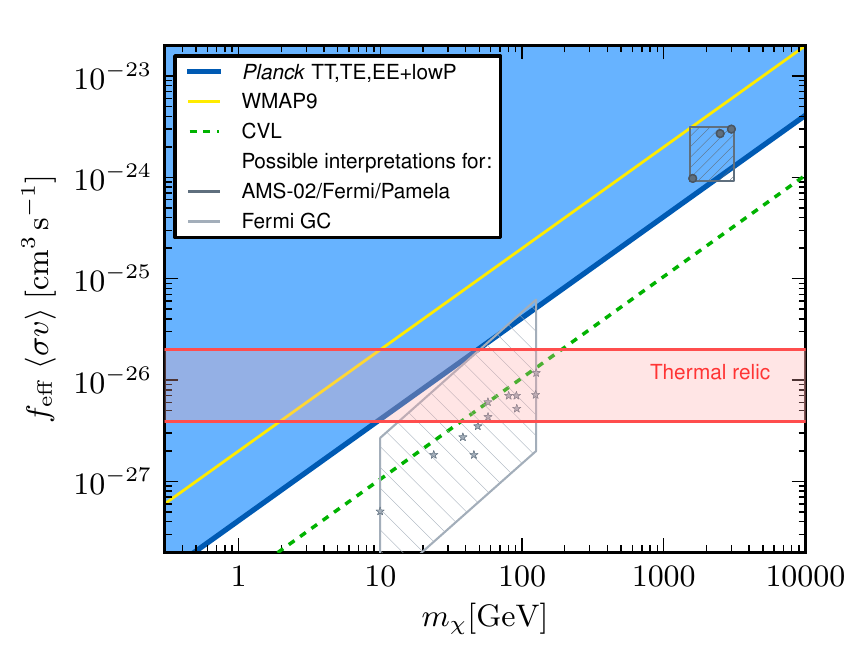}
  \caption{Left: Fermi-LAT 95\% confidence level constraints on the dark-matter
    annihilation cross section to $b\bar b$,\protect\cite{Ackermann:2015zua} compared to a
    calculation of the thermal relic cross
    section.\protect\cite{Steigman:2012nb}
    Right: Planck limits on the dark-matter annihilation cross section
    as obtained from the structure of CMB.\protect\cite{Ade:2015xua}
  }
  \label{fig:fermi-indirect-DM-limits}
\end{figure}

Strigari\,\cite{StrigariProceedings} reviewed recent results on indirect
dark matter detection.
In particular, the Fermi-LAT collaboration has combined upper limits
on gamma ray rates from Milky Way dwarf spheroidal satellite galaxies
(dSphs) with information about the satellites' dark matter mass, in
order to place constraints on the annihilation cross section for
various channels.
The constraints on one specific channel, $\chi\chi \to b\bar b$, are
illustrated in Fig.~\ref{fig:fermi-indirect-DM-limits}.
They are compared to the cross section that is required to obtain the
right thermal relic density and one sees that the resulting limit on
the DM particle mass, $m_\chi \gtrsim 100\GeV$, is in the same
ballpark as scales that are being probed at the LHC.

Limits on annihilation cross sections are also being placed by Planck
data on the CMB, since annihilation products would inject energy into the
gaseous background, modifying the CMB peaks.
Those limits, shown in Fig.~\ref{fig:fermi-indirect-DM-limits} (right)
depend only moderately on the annihilation channel, through a factor
$f_\text{eff}$ that encodes the fraction of rest-mass energy that is
injected into the gaseous background. 
The resulting lower limits on the mass of the dark-matter particle are
at the level of a few tens of GeV.

\begin{figure}
  \centering
  \begin{minipage}{0.48\linewidth}
    \includegraphics[width=\textwidth]{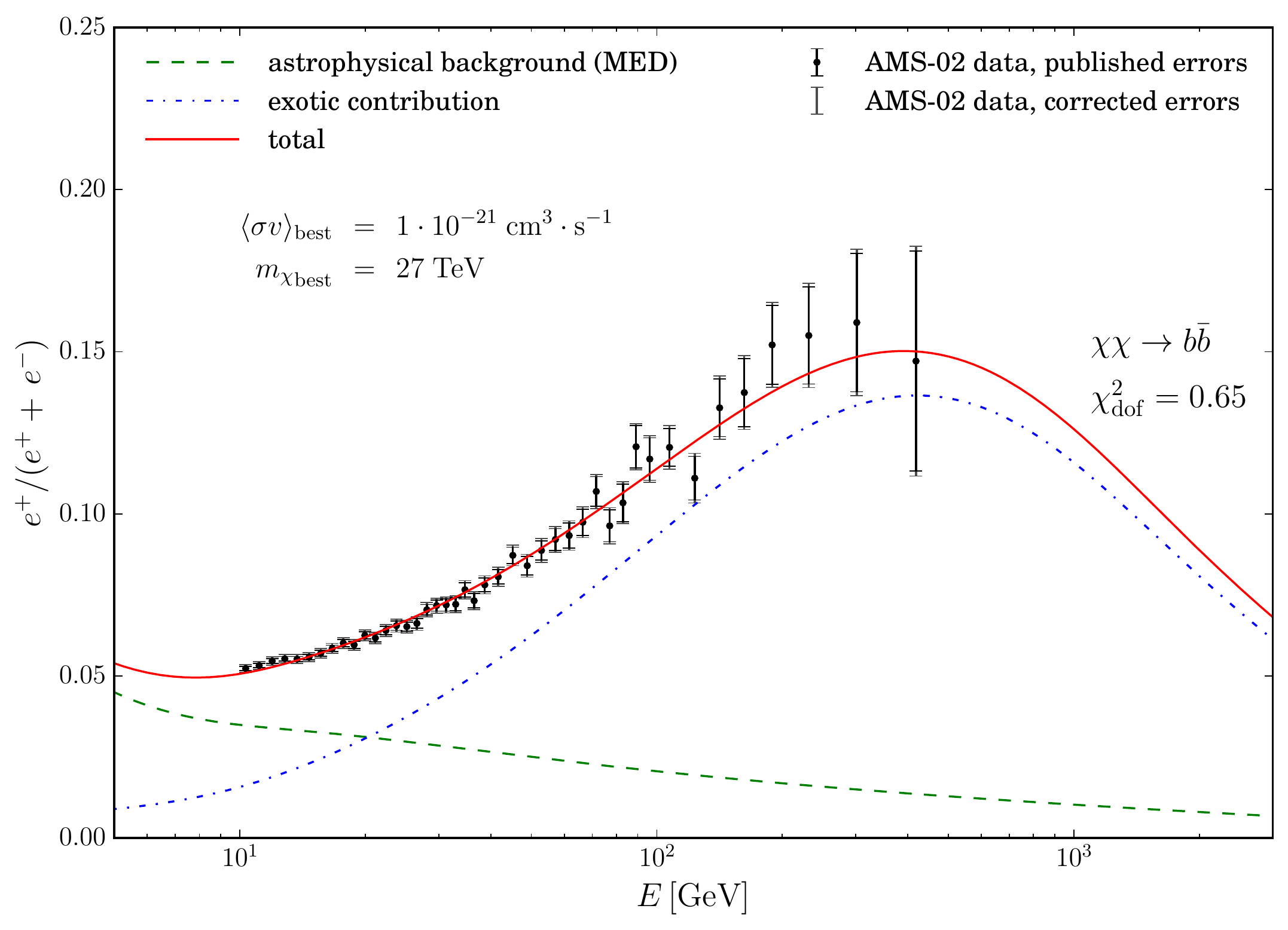}
  \end{minipage}\hfill
  \begin{minipage}{0.48\linewidth}
    \includegraphics[width=\textwidth]{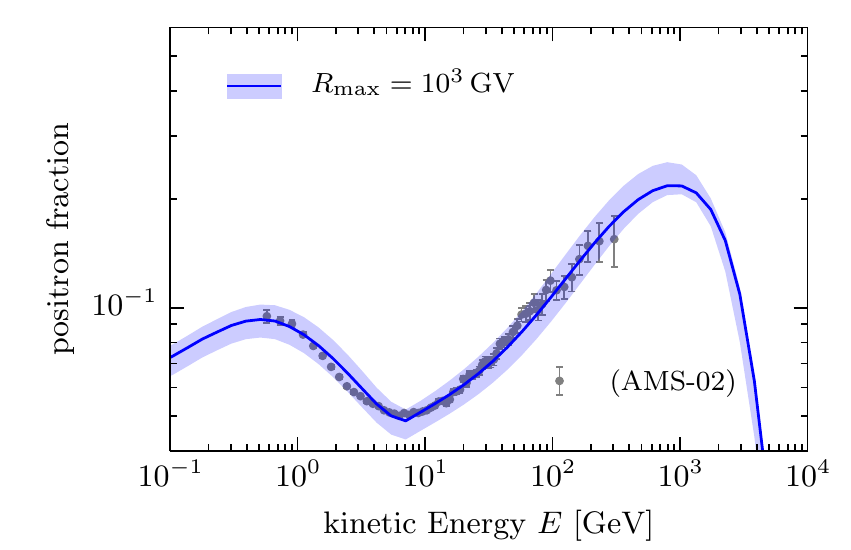}
  \end{minipage}  
  \caption{Left: fit to the AMS data on the positron fraction, shown
    as a function of energy, including a component from dark matter
    annihilation, $\chi\chi \to b\bar b$.
    Right comparison of the same AMS data to a model for production
    and acceleration of cosmic rays in supernova remnants.
    Figures as shown by
    Derome,\,\protect\cite{DeromeProceedings}
    taken from Refs.~\protect\cite{Boudaud:2014dta,Mertsch:2014poa} 
  }
  \label{fig:CR-positrons}
\end{figure}

One source of potential hints about dark matter in recent years has
come from an observed increase in the fraction of positrons relative to
electrons with increasing energy and also of the fraction of
anti-protons relative to protons.
Such increases would be expected at energies in the vicinity of the
mass of dark-matter particle.
The increases have been confirmed by recently released data from the
AMS experiment.\cite{DeromeProceedings}
However there can also be astrophysical explanations (e.g., in the
case of positrons, involving pulsars and supernova remnants) for such
an increase, and it appears to be difficult to distinguish between
different explanations, cf.~Fig.~\ref{fig:CR-positrons}.

\begin{figure}
  \centering
  \includegraphics[width=0.6\textwidth]{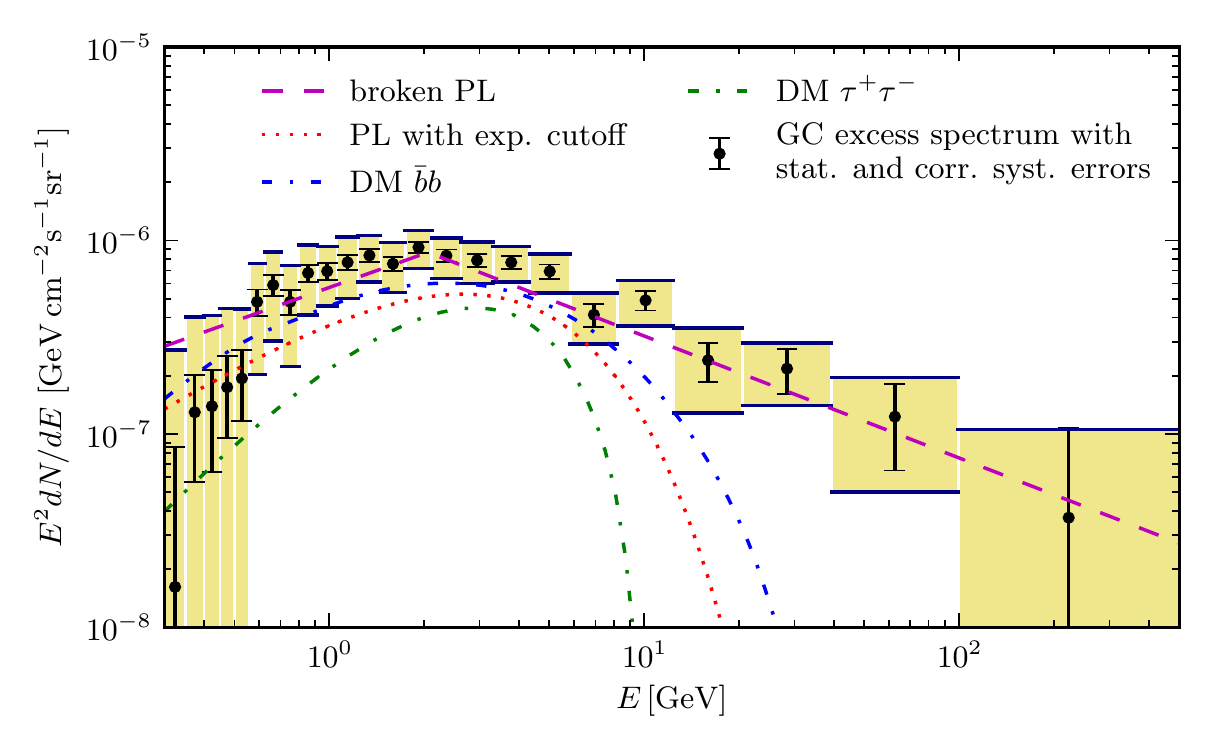} 
  \caption{The excess in the galactic-centre $\gamma$-ray spectrum,
    measured by the Fermi experiment,
    with comparisons to spectra from illustrative dark-matter models
    ($\chi\chi \to b\bar b$ with $m_\chi = 49\GeV$ and
    $\chi\chi \to \tau^+\tau^-$ with $m_\chi = 10\GeV$) and
    modifications of simple power laws
    (PL).\protect\cite{Calore:2014xka}
    Figure as shown in the talk by Linden.\,\protect\cite{LindenProceedings}
  }
  \label{fig:galactic-center-excess}
\end{figure}

One excess that has yet to find well-fitting astrophysical
explanations was discussed by Linden.\cite{LindenProceedings}
This is an excess in gamma-rays, peaked around 2~GeV, originating from
the galactic centre (but also up to $10^\circ$ away), with the spectrum
shown in Fig.~\ref{fig:galactic-center-excess}.
It was stated that it is very resilient to changes in the background
modelling, for example with a spectrum harder than expected for
astrophysical pion emission.
Dark-matter interpretations generally involve particle masses in the
range of $10-40\GeV$.\cite{CerdenoProceedings}
One of the key questions for future observations will be whether this
excess is found to be present also in other regions of dark matter
concentration, notably in dwarf galaxies.

The ideal indirect dark-matter detection signal would of course be an
excess of gamma rays at a very specific energy, i.e.\ a sharp line
feature.
One such feature had been claimed around
$133\GeV$,\,\cite{Weniger:2012tx} however with the most recent Fermi-LAT
data the signal appears to have largely
disappeared.\,\cite{TheFermi-LAT:2015gja}

\begin{figure}
  \centering
  \begin{minipage}{0.48\linewidth}
    \includegraphics[width=\textwidth]{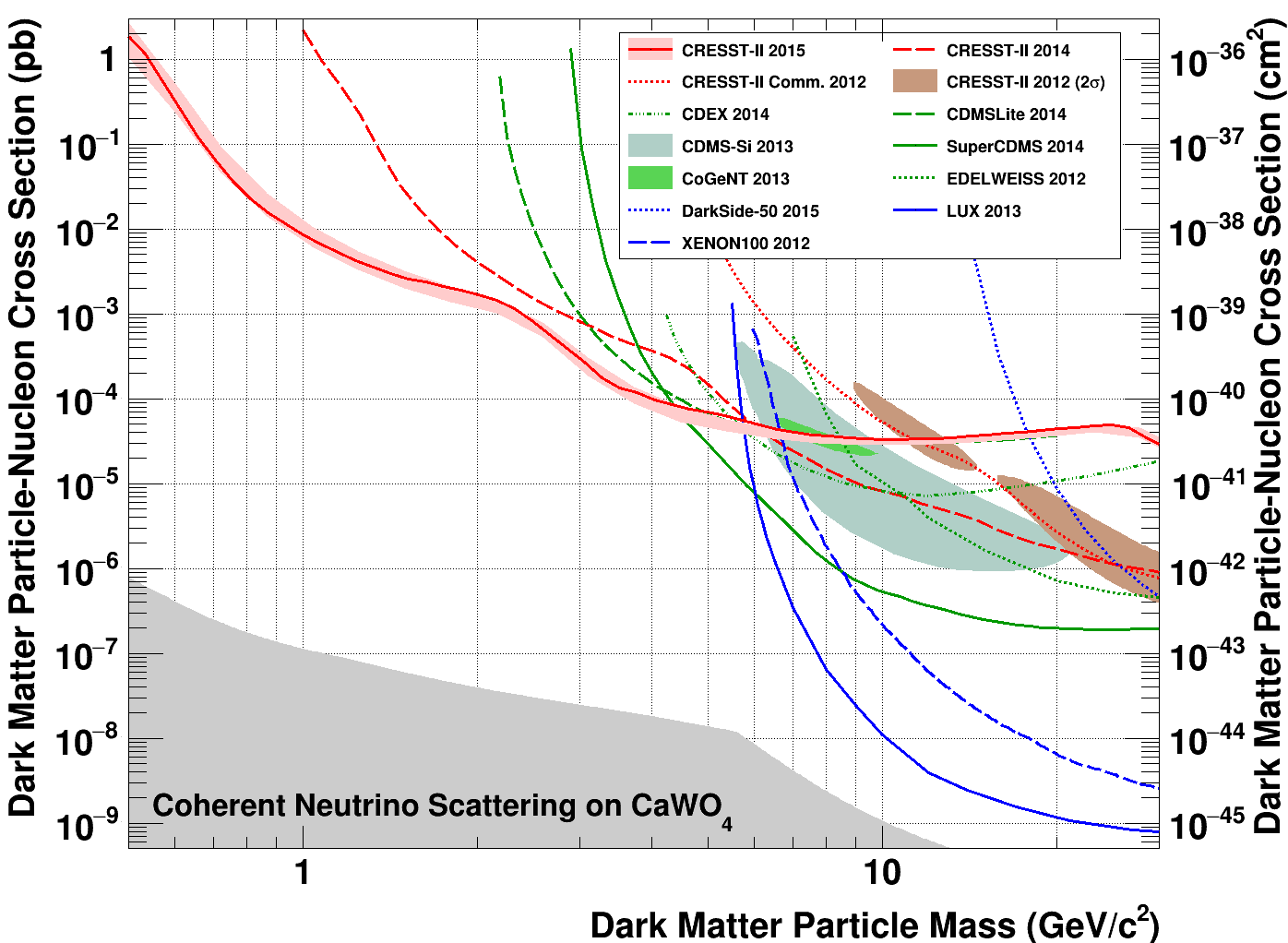}
  \end{minipage}\hfill
  \begin{minipage}{0.48\linewidth}
    \includegraphics[width=\textwidth]{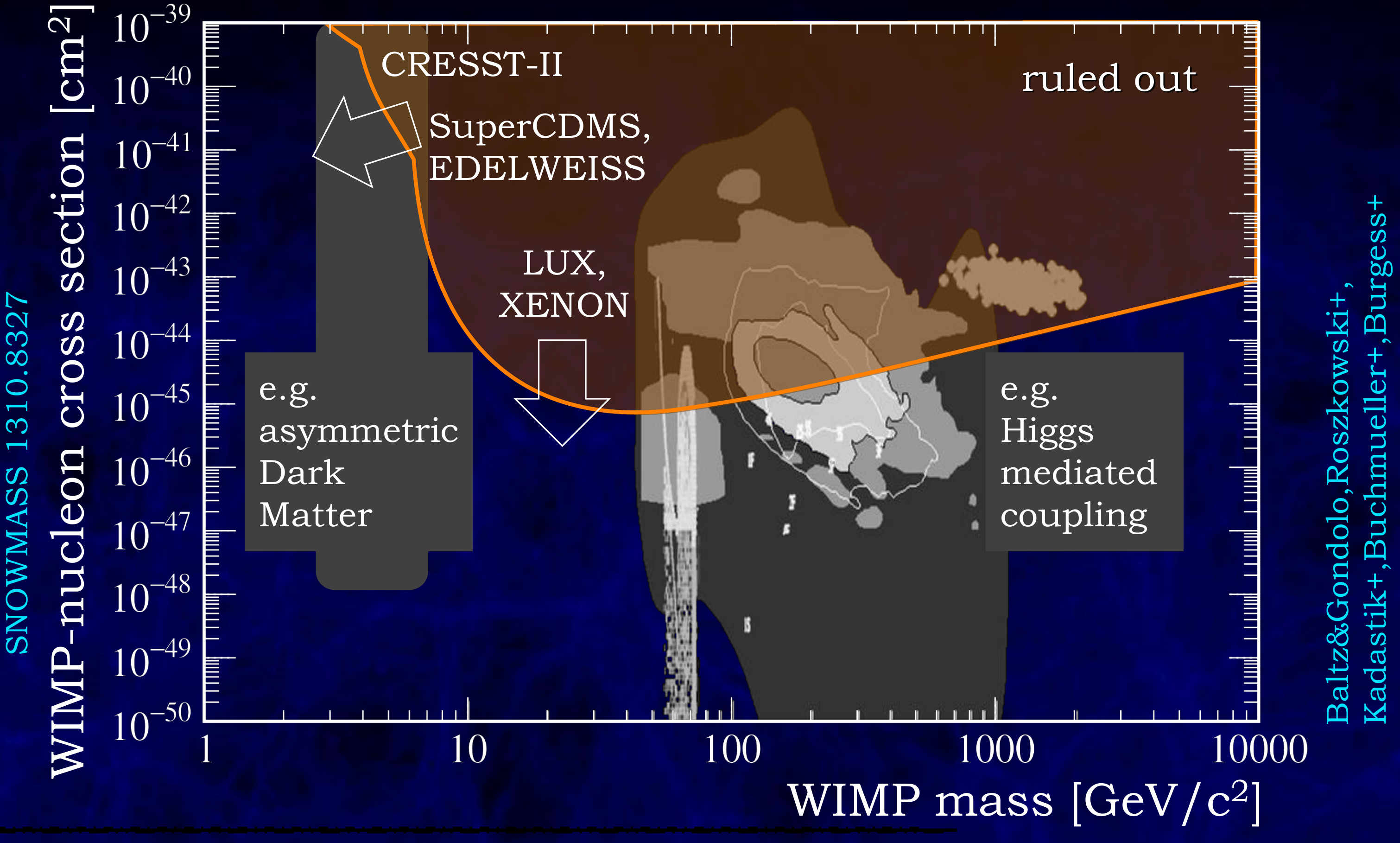}
  \end{minipage}
  \caption{Left: current limits from direct dark matter
    searches, as shown in the talk by
    Cerde\~no.\protect\cite{CerdenoProceedings} (Figure taken from
    Ref.~\protect\cite{Angloher:2015ewa}). 
    Right: comparison of today's limits with the dark-matter mass and
    nucleon coupling from the parameter space of various
    models. (Figure taken from Lang's talk.\,\protect\cite{LangProceedings})
  }
  \label{fig:direct-dark-matter}
\end{figure}

\begin{figure}
  \centering
  \includegraphics[width=0.7\textwidth]{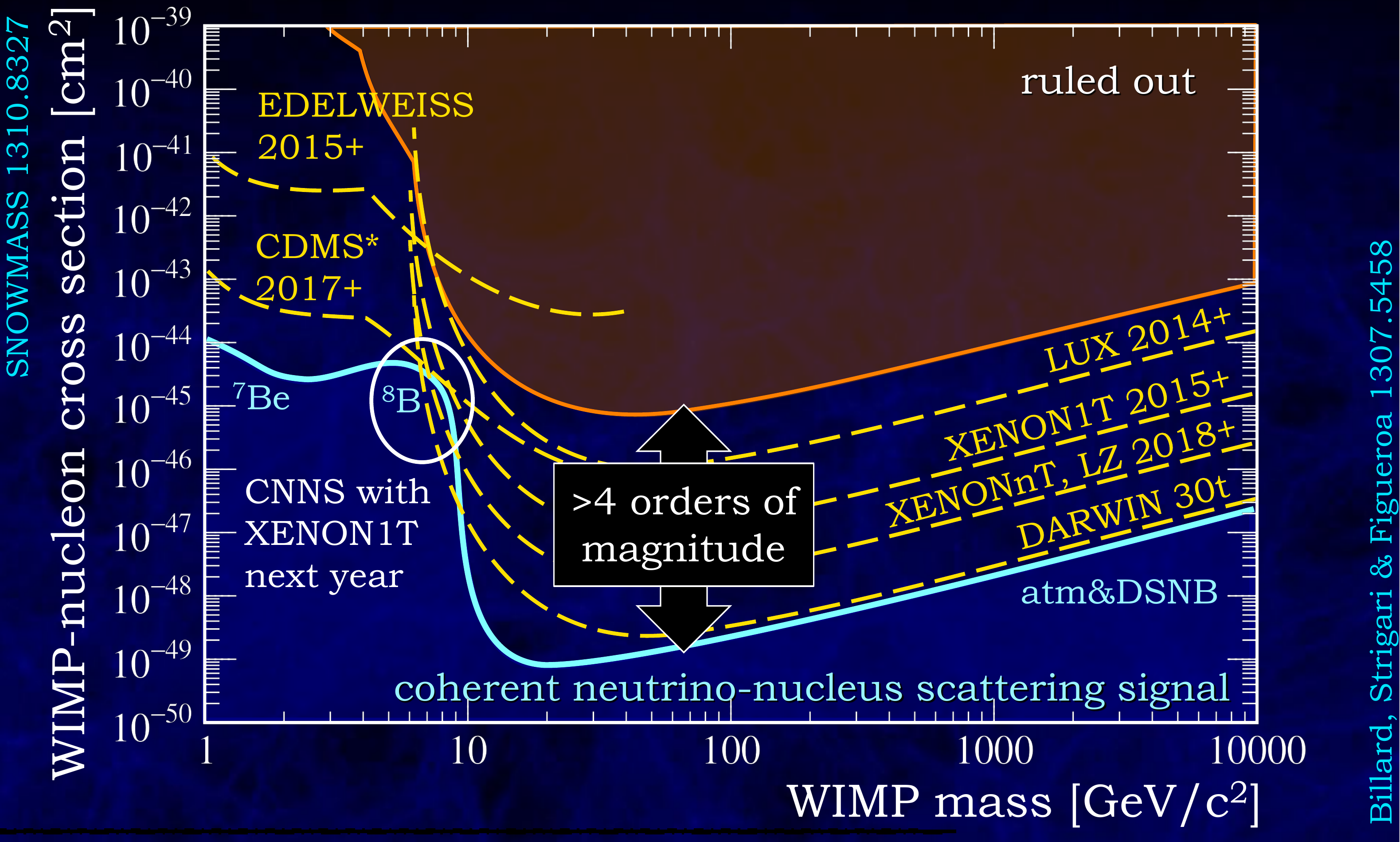}
  \caption{Prospects for dark-matter searches as shown in the talk by
    Lang.\protect\cite{LangProceedings}}
  \label{fig:DM-prospects}
\end{figure}

Dark matter is also being searched for through direct detection
experiments, which most commonly look for evidence of nuclei recoiling
after a coherent elastic interaction with a dark-matter particle.
While there have been various excesses over the years, at least one of
which still remains unexplained, overall the picture is largely one of
exclusion limits. 
The current status of the limits for the WIMP-nucleon cross section
v.\ WIMP mass is shown in Fig.~\ref{fig:direct-dark-matter}, including
also a plot of expected cross section and mass values in a range of
models.
One should be aware that such a figure comes with numerous
assumptions: that of an isothermal spherical halo, dark matter with
only spin-independent interactions, a coupling to protons that is
similar to that to neutrons, and that the scattering would be
elastic.\cite{CerdenoProceedings}
Within these assumptions, progress has been remarkable, with
Lang\,\cite{LangProceedings} pointing out that cross-section sensitivity
has been doubling every year for the past several years. 
Rapid progress is expected to continue for a number of years still:
for masses above $10\GeV$, there are about 4 orders of magnitude in
cross section between current limits and the coherent neutrino-nucleus
scattering signal expected from atmospheric neutrinos and the diffuse
supernova neutrino background,
and there is a well-defined programme of experiments that should
eventually be able to reach that limit.
There are also impressive prospects for progress in sensitivity to
masses in the few-GeV range, with many orders of magnitude improvement
in cross-section sensitivity expected from Edelweiss's 2015 data and
CDMS around 2017.
Meanwhile, in the region of sensitivity to masses around $10\GeV$,
where the $^8$B solar-neutrino background is particularly strong, one
may expect observation of those neutrinos in XENON1T already
next year.

%

One interesting point of comparison between LHC physics and direct
dark-matter detection (aside from the fact that they may both look for
the same models) is the increasing use of effective theories: these
are under much discussion for constraining Higgs
properties\,\cite{MassoProceedings} and are being examined also for dark
matter searches, as discussed by Cerde\~no,\,\cite{CerdenoProceedings} where they help to
systematically identify the very important complementarity between
different detector materials in terms of sensitivity to different
operators.
Such information may play an important role in guiding
the design of future direct detection experiments.

\section{Cosmology}

There are many questions that cosmology may help us solve: for
particle physicists, it can bring insight into questions such as dark
matter annihilation or neutrino masses, and it of course also brings
insight into questions that are more directly cosmological, e.g.\ the
fundamental characteristics of inflation and dark energy.

\begin{figure}
  \centering
  \includegraphics[width=0.6\textwidth]{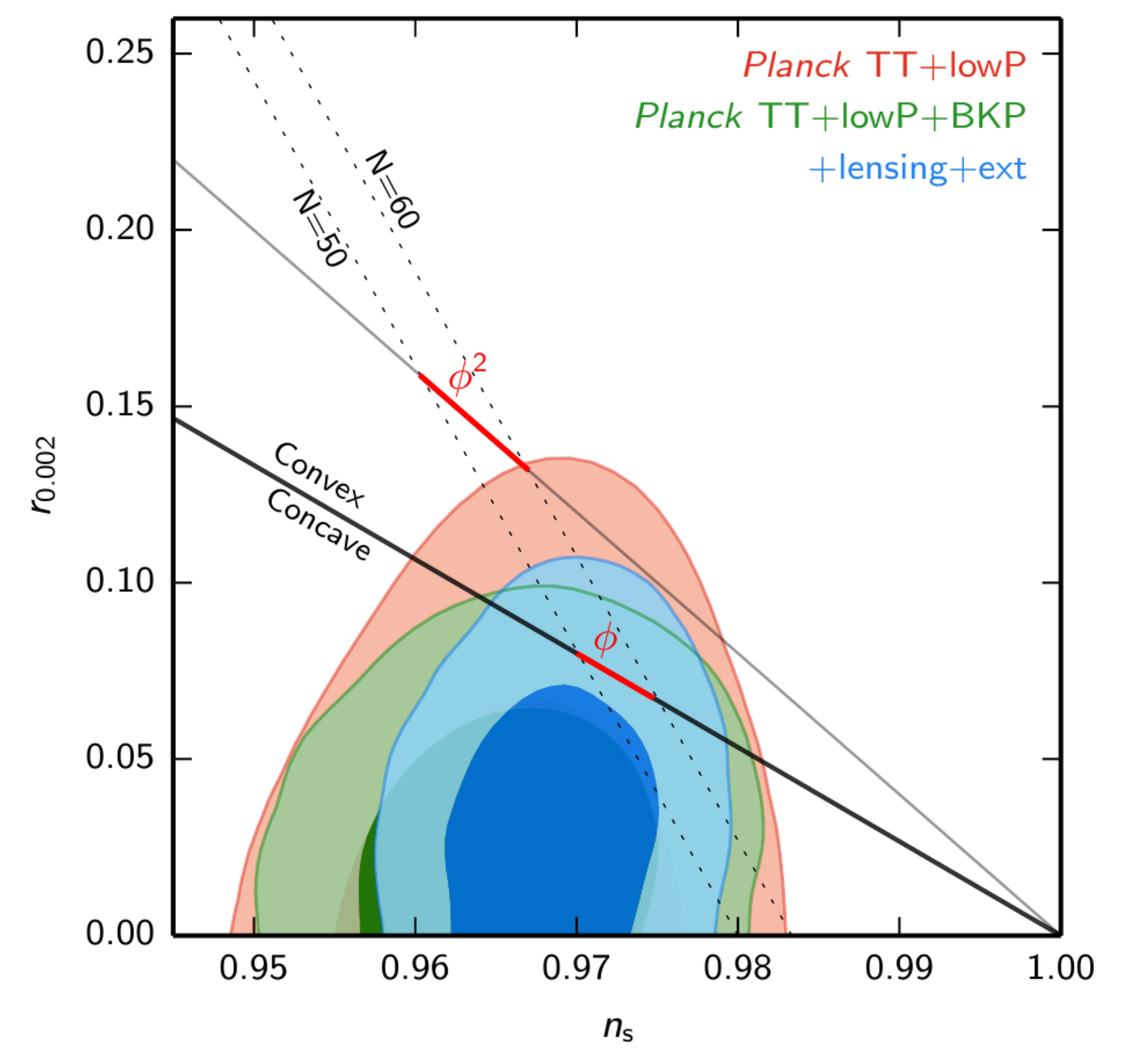}
  \caption{Results on the limit of tensor to scalar perturbations from
    the joint analysis of Planck and BICEP-Keck data, shown as a
    function of the scalar perturbation spectral index $n_s$ (taken
    from the talk by Ensslin\protect\cite{EnsslinProceedings}).}
  \label{fig:planck-bicep-keck}
\end{figure}

Ensslin's talk\,\cite{EnsslinProceedings} summarised some of the amazing
array of results from Planck.
One result that had been particularly awaited was the joint Planck and
BICEP/Keck analysis of the ratio of tensor to scalar perturbations,
$r$.
The reader almost certainly does not need reminding about the
excitement over BICEP/Keck's earlier apparent observation of a
non-zero $r$, which offered the hope of bringing detailed insight into
some of the physics at play in inflation.
The latest analysis involved a more robust separation of three
components: the intrinsic tensor perturbations, the contributions from
dust and those from synchrotron radiation.
Ultimately, as is now well known, the data no longer point to the
presence of tensor perturbations, but instead just place an upper
limit on their magnitude, as illustrated in Fig.~\ref{fig:planck-bicep-keck}.
Future prospects for improvements were discussed in the talk by
Ahmed.\cite{AhmedProceedings} 

\begin{figure}
  \centering
  \includegraphics[width=0.6\textwidth]{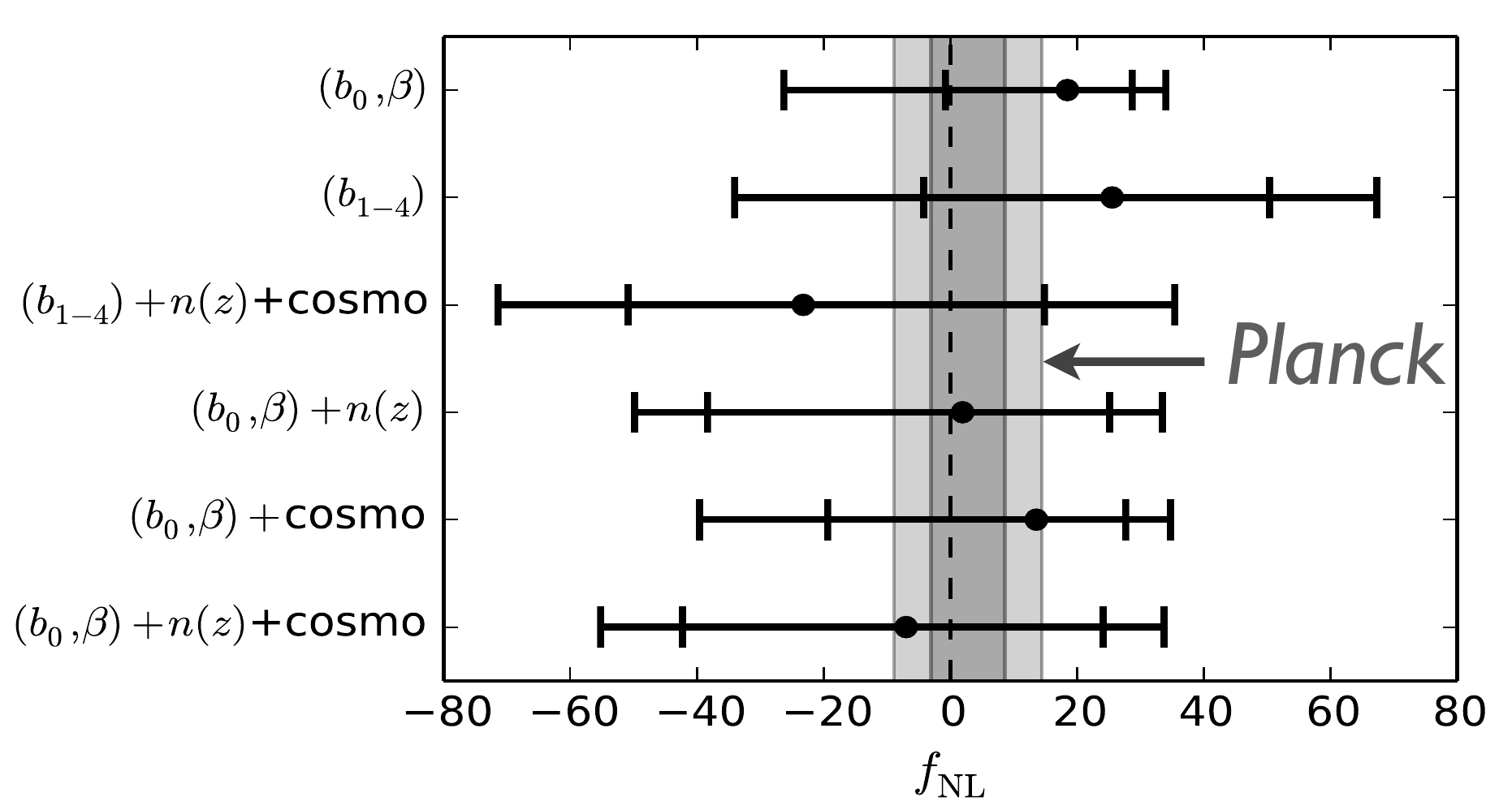}
  \caption{Limits on the primordial non-Gaussianity parameter
    $f_\text{NL}$ as obtained from a variety of methods  (taken from the
    talk by Peiris\,\protect\cite{PeirisProceedings}).}
  \label{fig:fNL}
\end{figure}

Another potential source of insight into the physics of inflation
would be the observation of primordial non-Gaussianity (PNG) in the spectrum
of scalar perturbations.
As discussed by Peiris,\cite{PeirisProceedings} one 
can obtain limits on PNG  from the CMB, for example from 3-point
correlations. 
One could also identify its impact on large scale structure,
specifically quasars, sensitive to PNG because it should lead to
enhanced clustering of massive objects.
Ultimately both methods indicate that any PNG is at best small, Fig.~\ref{fig:fNL}.

\begin{figure}
  \centering
  \includegraphics[width=0.6\textwidth]{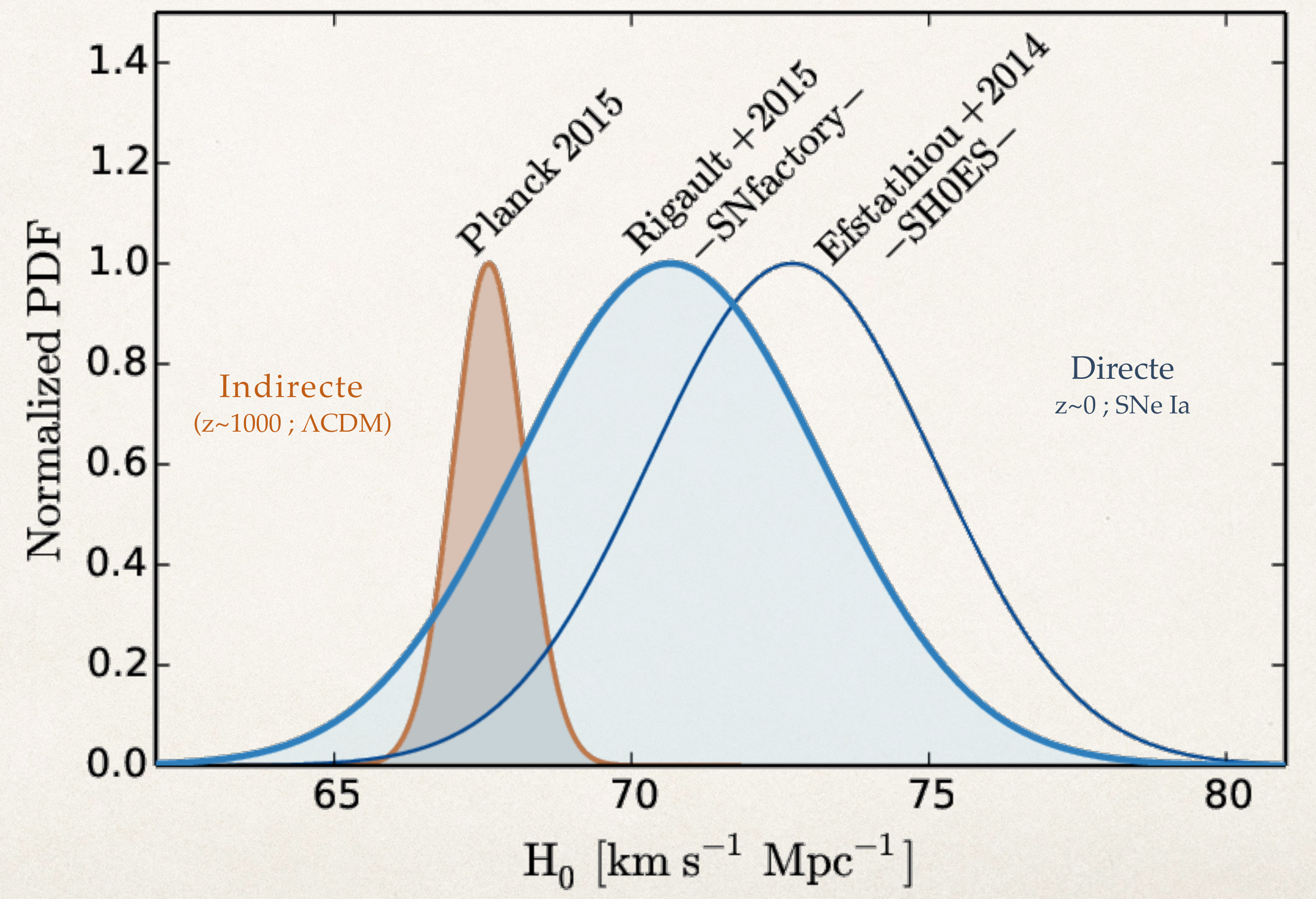}
  \caption{Extractions of the Hubble constant from Planck and from two
    supernova studies, one of which (SNfactory) takes into account the
    impact of environment on the supernova brightness (taken from the
    talk by Rigault\,\protect\cite{RigaultProceedings}).}
  \label{fig:Hubble}
\end{figure}

While the questions of tensor fluctuations and PNG bring may insight into
the early universe, another pressing question is that of dark energy,
which appears to dominate today's universe. 
As discussed in Rigault's talk,\cite{RigaultProceedings} it had
appeared that there was tension at the $2.5\sigma$ level between
extractions of the Hubble constant from Planck data and from
supernovae. 
Supernovae are useful because the uniformity of their
brightness makes it possible to use them to estimate distances.
Rigault presented new results that indicate that type 1A supernovae in
star-forming environments are somewhat fainter than other type 1A
supernovae. 
This helps resolve the tension, as illustrated in Fig.~\ref{fig:Hubble}.

\section{Concluding remarks}

There is some palpable frustration in the fields both of particle
physics and cosmology at the lack of confirmed discoveries of physics
beyond their respective standard models. 
This is despite the existence of fundamental open problems, such as the
identification of dark matter, gaining insight into the hierarchy
between the electroweak and Planck scales,
probing the nature of dark energy or 
understanding 
the
origin of the baryon asymmetry of the universe.

Nevertheless, there is amazing progress in improving experimental
sensitivity to new phenomena, as well as in the theory tools that help
us interpret the experimental results.
%
In tandem with this progress,
we are substantially expanding our knowledge about
cosmological and particle physics parameters, including Higgs
and neutrino properties.
%
One should also keep in mind
the long-term importance of today's
many null searches: in the future when something is discovered, it
will in part be because of those null searches that we may be able to narrow
down the viable candidates for explaining the discovery.

Even without immediate breakthroughs, there remains much to be learnt
and probed about our universe, and it is through that effort of
probing, in the broadest range of ways and making the best of the tools that
we can design, that we will ultimately be in a position to make
whatever discoveries Nature places within our reach.

\section*{Acknowledgments}

I wish to thank the organizers for the invitation to give this summary
talk and also for financial support to attend the conference.
I benefited much from conversations with many of the conference
participants, and would especially like to thank William Barter,
Jacques Dumarchez, Tim
Gershon, Concha Gonzalez Garcia, Uli Haisch, Raphael Lang, and Alison
Lister for sharing their insights, as well as Alessandra Tonazzo and
Andi Weiler for reading and commenting
on the manuscript.
Any innacuracies or misunderstandings are, however, of course entirely
my own.

\end{document}